\documentclass[sigplan,screen]{acmart}

\AtBeginDocument{%
\providecommand\BibTeX{{%
\normalfont B\kern-0.5em{\scshape i\kern-0.25em b}\kern-0.8em\TeX}}}

\usepackage{tikz}
\usepackage{algorithm}
\usepackage{algorithmic}
\usepackage{graphicx}
\usepackage{textcomp}
\usepackage{xcolor}
\usepackage{multirow}
\usepackage{hyperref}
\usepackage{thumbpdf}
\usepackage{booktabs}
\usepackage{colortbl}
\usepackage{url}
\usepackage{array}
\usepackage[normalem]{ulem}
\usepackage{xspace}
\usepackage{wrapfig}
\usepackage{tcolorbox}
\usepackage[framemethod=tikz]{mdframed}
\usepackage{enumitem}
\usepackage{balance}
\usepackage{tabularx}
\usepackage[justification=centering]{caption}
\usepackage{subcaption}
\usepackage[utf8]{inputenc}
\usepackage{siunitx}
\usepackage{pifont}
\usepackage{stackengine}

\def\ie{\textit{i.e.},~}
\def\etal{\textit{et al.}~}

\def\Snospace~{\S{}}

\newcommand{\para}[1]{\smallskip\noindent {\bf #1} }

\newcommand{\name}{\textsc{RCACopilot}\xspace}
\newcommand{\company}{Microsoft}
\newcommand{\team}{Transport team}
\newcommand{\service}{Transport}

\newcommand{\point}[1]{\vspace{1mm}\noindent\textbf{#1}.}



\begin{document}
\acmYear{2024}\copyrightyear{2024}
\setcopyright{acmlicensed}
\acmConference[EuroSys '24]{European Conference on Computer Systems}{April 22--25, 2024}{Athens, Greece}
\acmBooktitle{European Conference on Computer Systems (EuroSys '24), April 22--25, 2024, Athens, Greece}
\acmPrice{15.00}
\acmDOI{10.1145/3627703.3629553}
\acmISBN{979-8-4007-0437-6/24/04}

\title{Automatic Root Cause Analysis via Large Language Models for Cloud Incidents}

\settopmatter{authorsperrow=1} 
\author{Yinfang Chen$^{\diamond\S}$, Huaibing Xie$^{\diamond\P}$, Minghua Ma$^{\triangle*}$, Yu Kang$^*$, Xin Gao$^*$, Liu Shi$^*$, Yunjie Cao$^*$\\ Xuedong Gao$^*$, Hao Fan$^*$,  Ming Wen$^\dag$,  Jun Zeng$^\ddag$, Supriyo Ghosh$^*$, Xuchao Zhang$^*$ \\Chaoyun Zhang$^*$, Qingwei Lin$^*$, Saravan Rajmohan$^*$, Dongmei Zhang$^*$, Tianyin Xu$^\S$}
 
\affiliation{%
\institution{
Microsoft$^*$, 
University of Illinois at Urbana-Champaign$^\S$,
Peking University$^\P$\\
Huazhong University of Science and Technology$^\dag$,
National University of Singapore$^\ddag$
}
\country{}
}

\renewcommand{\authors}{Yinfang Chen, Huaibing Xie, Minghua Ma, Yu Kang, Xin Gao, Liu Shi, Yunjie Cao, Xuedong Gao, Hao Fan, Ming Wen, Jun Zeng, Supriyo Ghosh, Xuchao Zhang, Chaoyun Zhang, Qingwei Lin, Saravan Rajmohan, Dongmei Zhang, and Tianyin Xu}
\renewcommand{\shortauthors}{Chen et al.}

\begin{abstract}
 
Ensuring the reliability and availability of cloud services 
necessitates efficient root cause analysis (RCA) for cloud incidents. 
Traditional RCA methods, which rely on manual investigations of data sources 
such as logs and traces, are often laborious, error-prone, 
and challenging for on-call engineers. 
In this paper, we introduce \name{}, 
an innovative \textit{on-call system}
\textit{empowered by the}
\textit{large language model} 
for automating RCA of cloud incidents. 
\name{} matches incoming incidents to corresponding incident handlers 
based on their alert types, 
aggregates the critical runtime diagnostic information, 
predicts the incident's root cause category, 
and provides an explanatory narrative. 
We evaluate \name{} using a real-world dataset 
consisting of a year's worth of incidents from \company{}.
Our evaluation demonstrates that \name{} achieves RCA accuracy up to 0.766.
Furthermore, the diagnostic information collection component of \name{} has been successfully in use at \company{} for over four years.

\end{abstract}

\thanks{$\diamond$
This research was primarily conducted during an internship at Microsoft Research Asia.
 
$\triangle$ Minghua Ma is the corresponding author.
}

\maketitle
 
\section{Introduction}
\label{sec:intro}
Cloud computing serves as an indispensable infrastructure for numerous applications 
and services upon which people rely daily. 
As the adoption of cloud services continues to grow, 
ensuring their reliability, availability, and security becomes increasingly vital~\cite{ResinOSDI2022,haozhe2023codec, ma2020diagnosing, chen2023rainmaker, wu2017automated, gu:sosp:23, sun:osdi:22, sun:osdi:20}. 
However, the complexity of cloud systems makes them vulnerable to a variety of incidents that could pose significant challenges to these crucial properties \cite{TraceArk}.
A typical incident life-cycle consists of four stages: (1) \emph{Detection} \cite{ma2021jump,zeng2021watson,zeng2022shadewatcher}: When an anomalous system behavior is observed, an alert is raised by monitors or users of the service (internal engineers or external customers). (2) \emph{Triaging} \cite{bansal2020decaf, chen2019empirical, chen2019continuous}: After the detection, the incident is assigned to the appropriate engineering team after an initial assessment. (3) \emph{Diagnosis} \cite{luo2014correlating}: Assigned on-call engineers (OCEs) inspect different aspects of the incident and have several rounds of back-and-forth communication to identify the root cause. (4) \emph{Mitigation} \cite{jiang2020mitigate, ahmed2023recommending}: Several actions are taken by OCEs to mitigate the incident and to restore service health.
 
Root cause analysis (RCA) is pivotal in promptly and effectively 
addressing these incidents. 
By accurately diagnosing the underlying problem and preventing its recurrence, 
RCA not only restores service availability swiftly 
but also fortifies the overall reliability of cloud services.
However, identifying the root causes of these incidents 
often represents a daunting and time-consuming task that 
requires significant human expertise and intervention \cite{ma2020diagnosing}.
 
Traditional approaches to cloud incident RCA 
typically involve the manual collection and analysis of various types of data, 
such as logs \cite{he2022empirical, zhang2023system, li2023did, liu2022uniparser, inam2022sok}, metrics \cite{ma2018robust, ganatra2023detection, yan2023aegis}, traces \cite{zhao2023robust, zeng2023traceark}, and incident tickets \cite{jiang2020mitigate, shetty2022autotsg}. 
This manual process is not only laborious and error-prone, 
but can also be challenging due to varying levels of available information 
- what we term as the ``information spectrum''. 
The ``information spectrum'' describes a continuum of information availability, 
ranging from situations with too little information to those inundated with an excess. 
At either end of this spectrum, RCA can become particularly challenging.
The relevant information for RCA might be buried within the voluminous data,
leading to an information overload for OCEs. 
OCEs may find it challenging to quickly pinpoint the relevant information amidst the sea of data, hindering efficient incident resolution. 
Conversely, OCEs also encounter situations 
where they lack the necessary information to understand 
and address the root causes of incidents accurately. 
Beyond these challenges, the collected data itself is often noisy, 
incomplete and inconsistent, further complicating the RCA process. 
 
Specifically, the engineering team documents the frequent troubleshooting steps 
in the form of troubleshooting guides (TSGs) 
to facilitate the handling of future incidents. 
However, the volume of TSGs is overwhelming for OCEs, 
making the search for the most relevant guide a time-consuming task 
that might cause system downtime. 
Moreover, TSGs struggle to keep pace with the ever-evolving nature of cloud systems, 
thus often falling short when new incident types emerge.  
Even when a relevant TSG is located, it may not cover all the intricacies of the specific incident.
This could be due to variations in system configurations, 
the presence of multiple interacting root causes, 
or previously unknown issues. 
 
At the heart of RCA lies the fundamental challenge of 
\textit{efficiently collecting and interpreting comprehensive, 
incident-specific data} within a limited time frame. 
OCEs must quickly discern the relevance of various data types to the incident at hand 
and interpret them correctly. 
However, the complexity and sheer volume of data generated by cloud systems 
often impede rapid decision-making.
Furthermore, the expertise required to analyze various data types, 
along with the diverse range of possible incident causes, 
exacerbates the difficulty of the task.
As a result, OCEs may spend an inordinate amount of time analyzing data 
and formulating hypotheses, detracting from time that could be better 
spent resolving the incident and restoring system functionality.
 
Data-driven and Artificial Intelligence (AI) techniques have been leveraged for 
automating the incident management
~\cite{chen2020incidental, chen2019continuous}.
While there are existing techniques that   
recommends relevant TSGs~\cite{jiang2020mitigate} 
and automates the workflows~\cite{shetty2022autotsg} of TSGs,
their utility is limited by the inherent challenges associated with TSGs.
Despite these automated processes, OCEs still find themselves investing significant manual effort 
in sifting through the vast amounts of information, 
interpreting the data, and identifying the root causes of incidents.
 
The recent advent and success of large language models (LLMs) in performing complex tasks~\cite{wei2022chain, lian2023configuration,jin2023assess}, suggests a promising avenue for enhancing RCA. 
Specifically, LLMs can be used to parse through high-volume data, discern relevant information, and produce succinct, insightful outputs. This significantly alleviates the burden on OCEs to manually sift through vast amounts of data, helping them focus on resolving the incident more quickly and effectively. Additionally, LLMs can adapt to new and evolving types of incidents, learning from previous data to improve future predictions. While LLMs can process and generate text efficiently, they lack intrinsic domain-specific knowledge, especially in specialized areas such as cloud incident management. This lack of understanding of specific contexts, such as cloud incidents, can limit their accuracy in predicting incident root causes and generating appropriate explanations. 
 
Recently, Ahmed \emph{et. al.} \cite{ahmed2023recommending} proposed to finetune a LLMs with domain-specific datasets for generating root causes of an incident just by leveraging the title and summary information available at the time of incident creation. While they have demonstrated promises of LLMs in incident root causing, finetuning has several limitations: 
(1) As accurate RCA requires various sources of complex unstructured or semi-structured data (e.g., logs, telemetry, traces, and natural language description), just using a generic title and summary  might miss useful signals to reach conclusive diagnosis details; 
(2) finetuning is costly and may require a huge volume of training samples;
(3) it is challenging to continuously update a finetuned GPT model with evolving nature and scope of incidents; therefore such models are prone to generate more hallucinated results over time.
 
In this paper, we introduce \name{}, a novel on-call system presenting an automatic end-to-end approach to cloud incident RCA. 
\name{} operates as an on-call system, empowering OCEs to construct `incident handlers' - automated workflows tailored to each alert type, made up of reusable actions reflecting OCEs' expertise. These predefined handlers automatically streamline the collection of incident-specific diagnostic information from multiple sources, thus ensuring a more focused and relevant data accumulation process to avoid issues on either end of the information spectrum. 
Subsequently, the LLM component of \name{} processes this diagnostic data, predicting the category label of incident root causes and providing corresponding explanations. The combination of incident handlers and the LLM allows \name{} to significantly enhance adaptability and scalability in incident response. As a result, \name{} can effectively handle a diverse types of incidents while reducing the need for extensive human intervention.

The diagnostic information collection component of \name{} has been in use at \company{} for over four years. 
In recent developments, a root cause prediction component was prototyped and, following a successful preliminary phase, has been actively deployed by an incident management team at \company{} for a period spanning several months.

\vspace{2pt}
\para{\bf Contributions.} This paper makes three main contributions:
\begin{itemize}[leftmargin=*]
\item We propose \name{}, an automated end-to-end solution 
for cloud incident root cause analysis 
that enables on-call engineers to construct incident-specific automatic workflows 
for efficient data collection from multiple sources.
\item We introduce the integration of a large language model within \name{} 
that autonomously analyzes the collected diagnostic data 
to predict incident root cause categories and generate explanations, 
demonstrating the potential of the large language model in root cause analysis.
\item We showcase the real-world applicability of \name{} 
by presenting its successful adoption within \company{}. 
This illustrates its practical effectiveness in enhancing root cause analysis efficiency, 
demonstrating the feasibility and benefits of our approach 
in real-world cloud scenarios.
\end{itemize}

\section{Background and Motivation}
\label{sec:background}
In this section, we first introduce the concept and importance of incident root cause analysis.
We then present real-world examples of troubleshooting guides and illustrate their inherent limitations.
Lastly, we discuss the potential advantages of integrating a large language model into the RCA process, which motivates our work.
 
\subsection{Incident Root Cause Analysis}
In the realm of cloud services, 
an incident refers to any event that disrupts normal service operations 
or causes degradation in the quality of services. 
When such incidents occur, root cause analysis is performed to 
identify the underlying issue causing the disruption.
 
RCA in cloud services is a multi-faceted process:
\begin{itemize}[leftmargin=*]
\item \textit{Data Collection:} Gathering incident-related data from various sources such as logs, metrics, traces, or alerts is the first step in RCA. 
\item \textit{Data Analysis:} The collected data is then analyzed to identify patterns, anomalies, or correlations that can possibly provide clues about the root cause of the incident.
\item \textit{Hypothesis Verification:} Based on the data analysis, hypotheses about the possible root cause are formulated and then verified by OCEs.
\end{itemize}
 
Given the complexity and dynamism nature of cloud systems, 
along with the immense volume of data involved, 
conducting RCA is a challenging task, which requires substantial expertise and time. 
Take the scale of \company{}'s email service as an example,
which delivers over 150 billion messages daily. 
Ensuring the smooth operation of such a large-scale service demands an efficient and effective RCA approach.
This is pivotal in maintaining a reliable and high-performing 
communication infrastructure, particularly for organizations 
that rely heavily on \company{}'s email servers.

\begin{figure}[t]
\centering
\begin{tcolorbox}[width=\linewidth, title={Troubleshooting Guide for Poisoned Messages}]
1. Go to the Poisoned Message Dashboard. This page gives a real-time, high-level view of the Poison Message feature. The charts should indicate whether the problem has resolved itself or is ongoing, as well as some sense of where it is occurring \dots{}\\
2. \textit{The Dashboard newly implements an Exception Table} that has poisoned messages within a time frame. In most cases, whatever exception is causing an alert will rise to the top of the table \dots{} \\
3. You may also check the Poison Message Logs \dots{}\\
\dots{}
\end{tcolorbox}  
\caption{
A TSG for a poisoned message incident.
}
\label{fig:tsg}
\end{figure}
 
\subsection{The Opportunities and Challenges of Multi-Source Data in Incident Management}
Managing incidents in the complex ecosystem of cloud services necessitates a comprehensive understanding of system states. This comprehension often stems from the consolidation of multi-source data, which includes traces, logs, and metrics. Traces represent tree-structured data detailing the flow of user requests, logs are semi-structured text recording hardware and software events, while metrics monitor service status or user-perceived metrics, forming time series data. While these individual data sources yield valuable insights, capitalizing on their potential has challenges. Traditional approaches such as TSGs, though useful, may fail to exploit the full wealth of multi-source data due to inherent limitations.
 
\subsubsection{Opportunities of Multi-Source Data}
Different data sources provide different perspectives on the system state. For instance, logs can offer detailed event sequences, metrics can reflect system performance over time, and traces can reveal the propagation of requests across services. Integrating these data sources can provide a more comprehensive view of the system, enabling more accurate and efficient incident diagnosis and resolution. Furthermore, multi-source data can facilitate correlation and causality analysis, which is crucial for root cause analysis. By analyzing the relationships between different data sources, we can identify patterns and anomalies that may indicate the root cause of an incident.
 
\subsubsection{Challenges of Multi-Source Data}
Despite its potential, effectively leveraging multi-source data in incident management is challenging. The sheer volume and complexity of data from various sources can be overwhelming, making it difficult to extract meaningful insights. Worse still, different data sources may provide inconsistent or conflicting information. Moreover, real-world data is often noisy, which can complicate analysis and lead to false conclusions.

\subsubsection{Limitations of TSGs}
Traditional TSGs represent an early attempt to leverage multi-source data for incident management. They guide OCEs to gather and analyze data from various sources to diagnose and resolve incidents. However, TSGs face several inherent limitations: 
\begin{itemize}[leftmargin=*]
\item \textit{Manual data integration:} TSGs typically require OCEs to gather data from different sources manually. This process can be time-consuming and error-prone. 
Notwithstanding the existence of diverse troubleshooting guides and TSG recommendation techniques~\cite{jiang2020mitigate}, dependence on TSGs still remains a significant stress and burnout for OCEs due to the inherent limitations of the manual process.
\item \textit{Outdated information:} TSGs, as static documents, often struggle to stay up-to-date with the evolving system changes and new insights about incident root causes. This lag can lead OCEs to follow outdated or suboptimal troubleshooting steps. For example, a new feature (``Exception Table'') to check Poison Message exceptions, mentioned as the second step in Figure~\ref{fig:tsg}, was not immediately incorporated into the TSG upon its release, causing potential inefficiencies in incident resolution.
\item \textit{Insufficient details and coverage:} High-level instructions often appear in TSGs, lacking in detail and specific guidance, which forces OCEs into additional research and prolongs incident mitigation. In the TSG example from Figure~\ref{fig:tsg}, the third step instructs to check the Poison Message Logs, 
leaving out crucial details and causing confusion for OCEs unfamiliar with this incident type. Additionally, TSGs may overlook common checks, e.g., disk space checks, leading to partial or inadequate incident resolutions.
\end{itemize}

\subsection{The Promise of Large Language Models for Incident Management}
The rapid advancements in natural language processing and machine learning 
have led to the development of powerful LLMs, 
which are reported to be effective at various downstream tasks 
with zero-shot and few-shot learning~\cite{brown2020language, chen2021evaluating, lian2023configuration}.
These models have shown exceptional performance in translation, summarization, and question-answering. 
Leveraging their potential for incident management in cloud computing systems could revolutionize the way OCEs identify and resolve incidents. By automating the interpretation aspect of incident management, LLMs can help alleviate the stress and cognitive load associated with complex on-call tasks for OCEs, 
which enables OCEs to focus more on higher-level jobs and decision-making.

\subsection{Our Motivation}
The motivation for our work is rooted in the challenges faced when using manual TSGs to diagnose incidents and identify the underlying root causes. 
Our goal is to develop an automated diagnostic process that harnesses the capabilities of LLMs to address various cloud incidents more effectively.
 
Different from previous work~\cite{shetty2022autotsg}, which employs AI techniques to generate automated workflow from existing TSGs, our goal is to enable experienced OCEs to construct an automated pipeline for incident diagnosis. This approach allows OCEs to be directly assisted in identifying the root cause without the need to investigate intermediate diagnostic information, though they still have the option to do so.
 
We envision a future in which root cause analysis is predominantly automated, requiring minimal manual verification only when necessary. Our approach seeks to provide OCEs with timely, relevant, and accurate information for specific incidents, leading to more efficient RCA. By leveraging LLMs to predict root cause category, our research aims to alleviate the stress and cognitive load associated with incident management, ultimately enhancing the efficiency and effectiveness in addressing incidents.

\begin{table*}[]
\caption{Examples of cloud incidents in different root cause categories.}
\vspace{-7.5pt}
\small
\begin{tabular}{c|cllclll}
\toprule
No. &
Sev. &
Scope &
Category &
Occur. &
Symptom &
Cause &
\\ \midrule
1 &
1 &
Forest &
\multicolumn{1}{p{3.8cm}}{AuthCertIssue} &
3 &
\multicolumn{1}{p{3.8cm}}{Tokens for requesting services were not able to be created. Several services reported users experiencing outages.} &
\multicolumn{1}{p{3.8cm}}{A previous invalid certificate overrided the existing one due to misconfiguration.} &
\\ \hline
2 & 
2 &
Machine &
HubPortExhaustion &
27 &
\multicolumn{1}{p{3.8cm}}{A single server failed to do DNS resolution for the incoming packages.} 
& \multicolumn{1}{p{3.8cm}}{The UDP hub ports on the machine had been run out.} &
\\ \hline
3 &
2 &
Forest &
DeliveryHang &
6 &
\multicolumn{1}{p{3.8cm}}{Mailbox delivery service hang for a long time.} &
\multicolumn{1}{p{3.8cm}}{Number of messages queued for mailbox delivery exceeded the limit.} &
\\ \hline
4 &
2 &
Forest &
CodeRegression &
15 &
\multicolumn{1}{p{3.8cm}}{An SMTP authentication component's availability dropped.} &
\multicolumn{1}{p{3.8cm}}{Bug in the code.} &
\\ \hline
5 &
2 &
Forest &
CertForBogusTenants &
11 &
\multicolumn{1}{p{3.8cm}}{The number of concurrent server connections exceeded a limit.} &
\multicolumn{1}{p{3.8cm}}{Spammers abused the system by creating a lot of bogus tenants with connectors using a certificate domain.} &
\\ \hline
6 &
1 &
Forest &
MaliciousAttack & 
2 &
\multicolumn{1}{p{3.8cm}}{Forest-wide processes crashed over threshold.} &
\multicolumn{1}{p{3.8cm}}{Active exploit was launched in remote PowerShell by serializing malicious binary blob.} &
\\ \hline
7 &
2 &
Forest &
UseRouteResolution &
9 &
\multicolumn{1}{p{3.8cm}}{Poisoned messages sent to the forest made the system unhealthy.} &
\multicolumn{1}{p{3.8cm}}{A configuration service was unable to update the settings leading to the crash.} &
\\ \hline
8 &
2 &
Forest &
FullDisk &
2 &
\multicolumn{1}{p{3.8cm}}{Many processes crashed and threw IO exceptions.} &
\multicolumn{1}{p{3.8cm}}{A specific disk was full.} &
\\ \hline
9 &
2 &
Forest &
InvalidJournaling &
11 &
\multicolumn{1}{p{3.8cm}}{Messages stuck in submission queue for a long time.} &
\multicolumn{1}{p{3.8cm}}{The customer set an invalid value for the \service{} config and caused TenantSettingsNotFoundException.} &
\\ \hline
10 &
3 &
Forest &
DispatcherTaskCancelled &
22 &
\multicolumn{1}{p{3.8cm}}{Normal priority messages across a forest had been queued in submission queues for a long time.} &
\multicolumn{1}{p{3.8cm}}{Network problem caused the authentication service to be unreachable.} &
\\ \bottomrule
\end{tabular}
\label{tbl:cloud_incidents}
\end{table*}

\section{Insights from Incidents}
\label{sec:study}
 
We conducted a comprehensive study of the one-year incidents from an email service from \company{}, employing rigorous qualitative analysis methods. Specifically, each incident was carefully reviewed and categorized based on the characteristics of the problem, the source of the issue, and the impact on the system by our experienced OCEs. We paid particular attention to the root causes of the incidents, the effectiveness of the response, and the recurrence of similar issues. While our insights were indeed intuitively derived, they were firmly grounded in empirical data and analysis. Our study not only yielded valuable insights into incident patterns and challenges but also informed the development and refinement of our approach.
 
\paragraph{\textbf{\textit{Insight 1: determining the root cause based on a single data source can be challenging}}}
\label{insight:1}
As an illustration, consider Incident 2 in Table~\ref{tbl:cloud_incidents}, where a single server failed to perform DNS resolution for incoming packets due to the exhaustion of UDP hub ports on a front door machine. This example highlights the difficulties in relying solely on a single source (monitor alert) to diagnose complex issues.
 
When a mailbox server sends mail to external email recipients, it uses specific front-door servers (proxies). However, each front-door server has a limited number of available SMTP outbound proxy connections. If a mailbox server's proxy connection request fails, it will be unable to send messages to external recipients. In this incident, the monitor first raises an alert indicating detected failures when connecting to the front door server. However, this alert only signifies a connection issue between the mail server and the front door server, without even suggesting a DNS resolution problem. Consequently, the root cause remains unclear.

\begin{figure}[h]
\centering
\includegraphics[width=0.35 \textwidth]{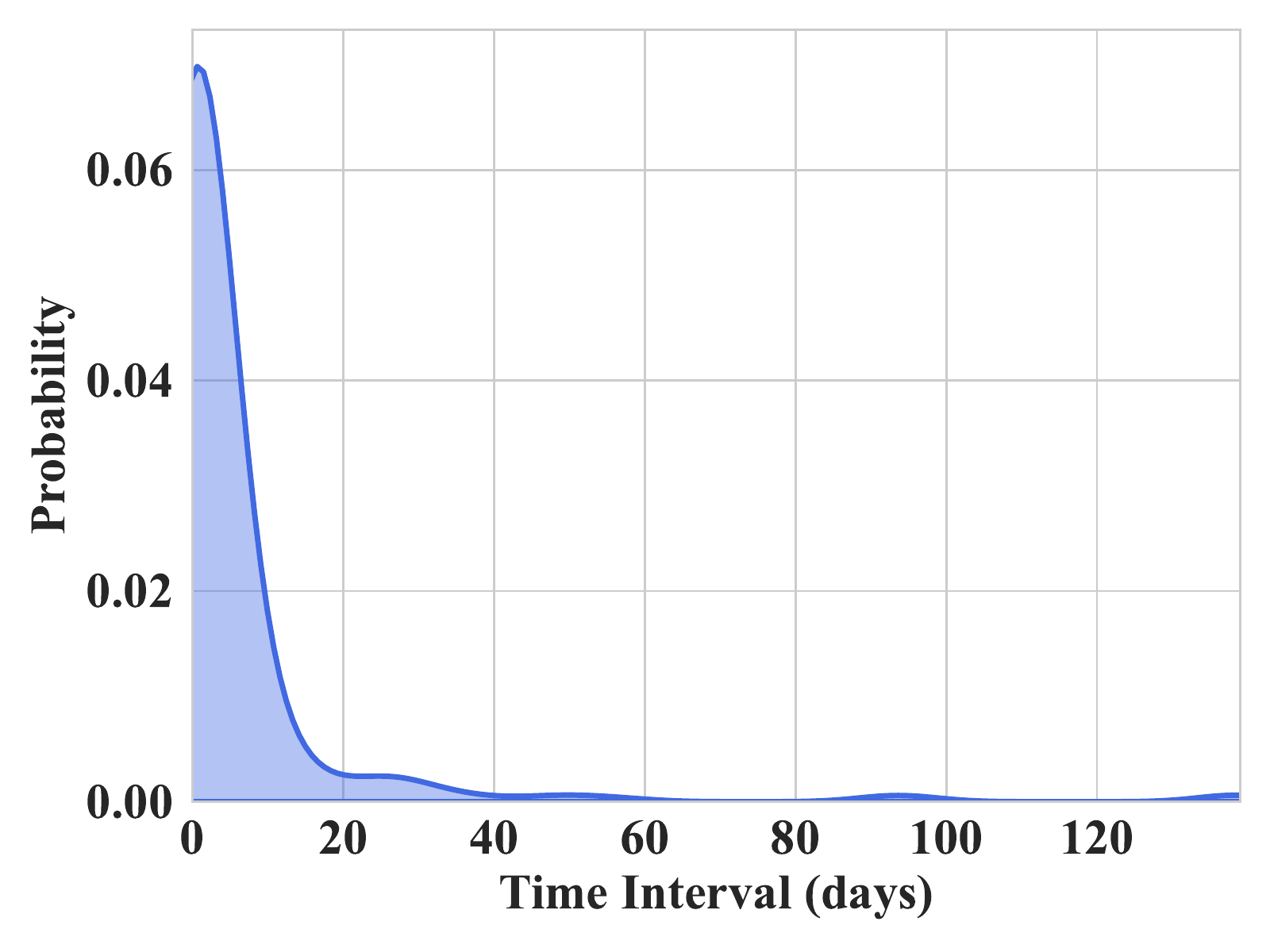}
\caption{
Recurring incidents proportion vs. time interval.
}
\label{fig:keyword_time_interval}
\end{figure}
 
\paragraph[short]{\textbf{\textit{Insight 2: incidents stemming from similar or identical root causes often recur within a short period}}}
\label{insight:2}
We found that most recurring incidents (93.80\%) tend 
to reappear within a brief span of 20 days,
as shown in Figure~\ref{fig:keyword_time_interval}. 
For instance, consider the category of Incident 9 from Table~\ref{tbl:cloud_incidents}. 
This type of incident, triggered by invalid customer configuration,
led to an accumulation of unprocessed messages in the queue, 
thereby significantly undermining its availability. 
Intriguingly, incidents of this category recurred 11 times in a span of merely 15 days.
Likewise, the DispatcherTaskCancelled incidents (No. 10 in Table~\ref{tbl:cloud_incidents}) 
and the DeliveryHang incidents (No. 3) reappeared 22 times and 6 times within a week 
and a single month, respectively.
These can be attributed to several factors. 
Unresolved root causes from the initial response may lead to the same issue re-emerging, 
especially if the problem is complex or not fully understood. 
Secondly, systemic vulnerabilities, if not addressed, can be repeatedly exploited, 
causing similar incidents. 
Thirdly, external dependencies, such as reliance on a service that frequently experiences outages, 
can also lead to recurring incidents. 
These patterns suggest that by leveraging insights from previous incidents, 
we could swiftly identify the root cause of new occurrences with the same root cause.

\begin{figure}[h]
\centering
\includegraphics[width=0.35 \textwidth]{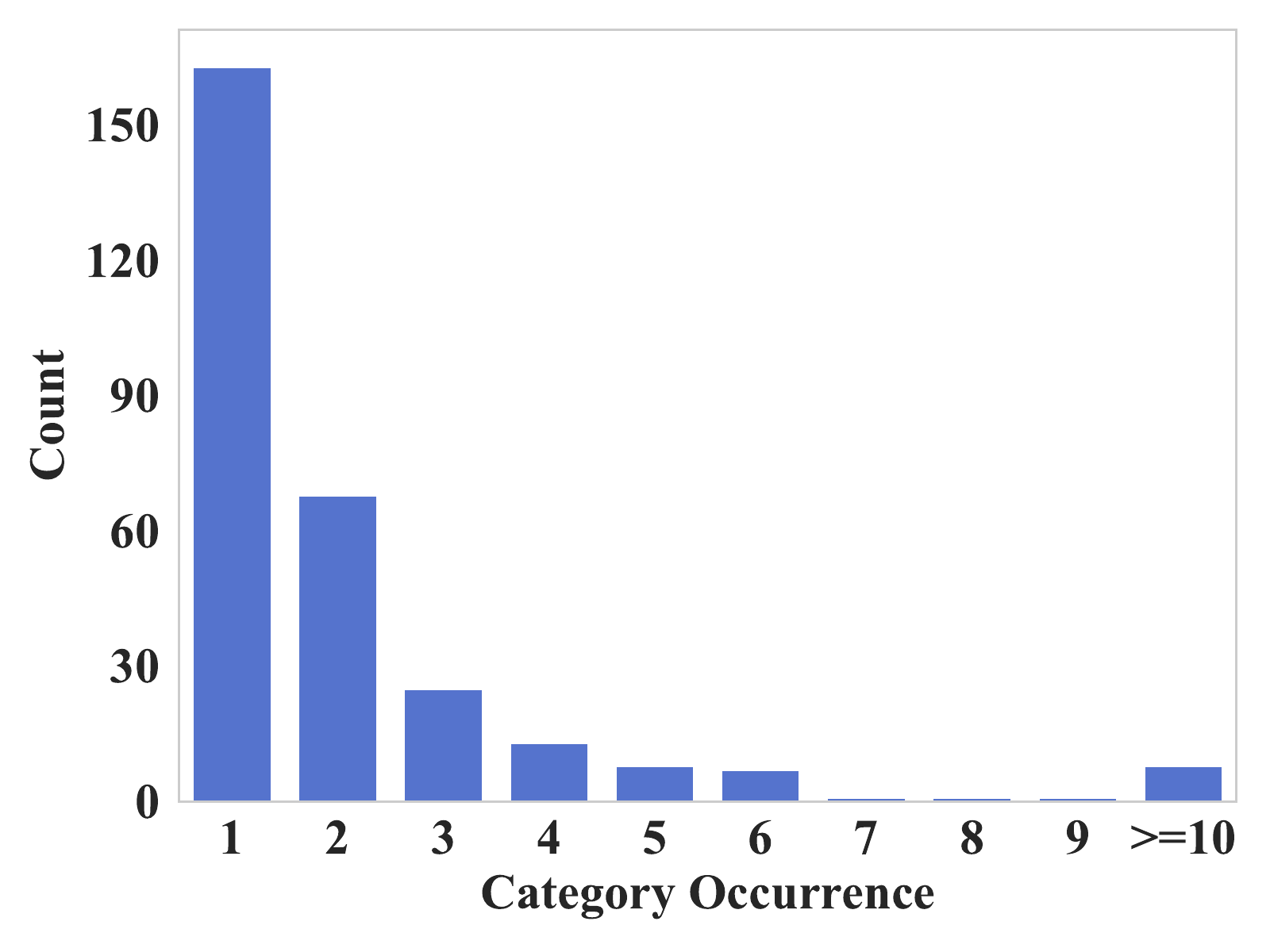}
\caption{
Distribution of incident category frequency.
}
\label{fig:incident_frequency}
\end{figure}
 
\paragraph[short]{\textbf{\textit{Insight 3: incidents with new root causes occur frequently and pose a greater challenge to analyze}}}
\label{insight:3}
TSGs can help OCEs diagnose issues 
by providing clear investigation guidance.
However, when incidents arise from new, previously unencountered root causes, 
OCEs face a set of challenges. 
For such incidents, no TSG exists, and OCEs may struggle to identify the underlying issues.
For instance, Incident 1 is a high-severity (severity 1) incident caused by misconfiguration,
which blocked the authentication token generation to lead to severe outages.
Similarly, Incident 6 is a malicious attack caused 
by an attacker launching an exploit with a malicious blob. 
This type of attack had never been encountered before, 
leaving OCEs without an existing TSG to reference. 
Lower severity level (severity 2) incidents, such as Incident 5, 
are also susceptible to this challenge when the spammer first abuses the system.
As Figure~\ref{fig:incident_frequency} shows, 
incidents with a new root cause category account for 24.96\% (163 among 653) of all incidents.
If OCEs spend their time searching for nonexistent TSGs,
the incident's impact could escalate further. 
Recognizing this challenge, it is necessary to propose a new approach 
that can effectively infer, categorize and explain the root causes for such unseen incidents, 
thereby reducing the time OCEs take to identify and address these unique incidents.

\section{\name}
\label{sec:approach}
 
\begin{figure*}[t]
\centering
\includegraphics[width=\textwidth]{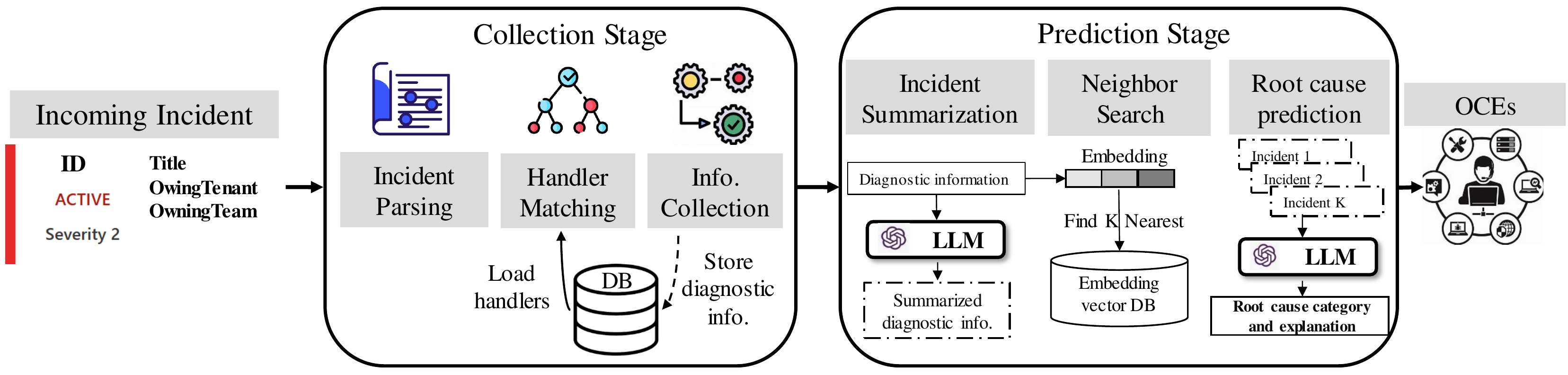}
\caption{
\name{} architecture.}
\label{fig:arch}
\end{figure*}
 
\name{} has two stages: the diagnostic information collection stage 
and the root cause prediction stage
as shown in Figure~\ref{fig:arch}.

\textbf{Diagnostic information collection stage}: This is the initial stage, where the incident is parsed and matched to the pre-defined incident handler. Each incident handler is tailored to a specific alert type. Upon matching the incident with the appropriate handler, \name{} proceeds to collect relevant diagnostic data from a variety of sources. 
 
\textbf{Root cause prediction stage}: Once the diagnostic information is collected, \name{} transitions into the root cause prediction stage. In this phase, \name{} applies its predictive module to determine the likely root cause category of the incident. This prediction is not a mere categorization, but it is also supplemented with an explanation detailing how \name{} arrived at the given prediction. Subsequently, the predicted category label is presented to experienced OCEs for review.

\subsection{Diagnostic Information Collection Stage}
\label{model-construct}
 
Driven by Insight-1 in Section~\ref{insight:1}, \name{} aims to collect multi-source data for RCA. 
Specifically, for each alert type, an incident handler is constructed, comprising a series of actions to collect diagnostic information. 
Alert types are used to categorize alerts based on specific monitors. 
Incidents sharing the same alert type exhibit similar symptoms, though they may stem from different root causes. 
 
The \name{} incident handler is a workflow that consists of a series of actions. Each action is a function that can be executed to collect specific diagnostic information from a target data source. OCEs can build and modify these handlers based on their expertise. The handler includes three distinct actions: \textit{scope switching action}, \textit{query action}, and \textit{mitigation action}, which will be explained in Section~\ref{sec:handler_action}.
Each action generates an output, guiding the control flow of the incident handler. We use a \name{} handler that diagnoses Incident 7 in Table~\ref{tbl:cloud_incidents} as an example to illustrate the handler usage.

\subsubsection{Incident handler}
The decision-making process that OCEs employ when handling an incident resembles a decision tree's control flow. The root node in the incident handler is the incident alert type, which is gathered from the system monitor.
We distilled OCE operations into three actions when constructing the incident handler.
As OCE operations can be similar to different incident types (e.g., conducting a common disk check or query to a database), we designed \name{} handler actions to be reusable across all handlers. 
We also maintain the versions of the handlers in the database, which can be used to track their historical changes.
 
\name{}'s incident handlers are constructed manually first and can be updated and modified dynamically by OCEs, allowing them to stay abreast with the most recent system changes and newly discovered root causes. For instance, when a new metric is introduced into the system, OCEs only need to construct a new action to collect the relevant data and incorporate it into the corresponding incident handler, which can ensure timely adaptation.

\subsubsection{Handler action} 
\label{sec:handler_action}
\begin{figure*}
\centering
\includegraphics[width=.8\textwidth]{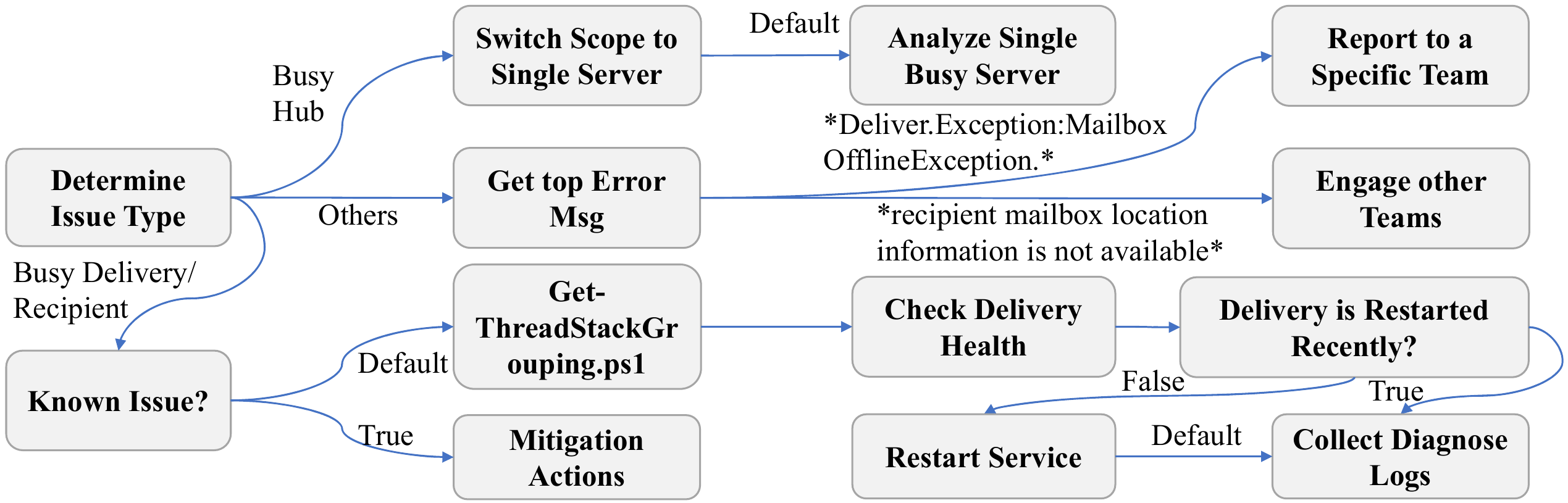}
\caption{
A \name{} handler for too many messages stuck in the delivery queue alert.
}
\label{fig:model}
\end{figure*}

\name{} leverages the synergy of multi-source data. The system uses predefined reusable actions in the incident handler to automatically collect relevant diagnostic information from diverse sources. The automated integration of data not only saves time but also reduces the likelihood of human error. It provides a more comprehensive view of the system state, facilitating efficient and accurate incident resolution. This significantly lightens the workload of OCEs, reducing stress and burnout, and enhancing the effectiveness of the incident resolution process. 
The action in the handler could be one of the following:

\textbf{Scope switching action}: This action facilitates precision in RCA by allowing adjustments to the data collection scope based on the specific needs of each incident. For instance, as depicted in Figure~\ref{fig:model}, if an alert originates at the `forest' level, signifying an issue within a specific forest, and the problem type is identified as `Busy Hub', the scope switching action can adjust the scope to the `machine' level. This modification allows for a more fine-grained investigation, specifically assessing if a singular hub server is overly taxed.
 
The implementation of this action ensures that we efficiently navigate the information spectrum. When the investigation requires a more targeted approach, this action can narrow the data collection scope. Conversely, if a more holistic view is necessary, it can widen the scope, say from a single machine to an entire forest. This flexibility contributes to a more balanced and effective diagnostic data collection.
 
\textbf{Query action}: Query action can query data from different sources and output the query result as a key-value pair table. 
This type of action can also be hooked to 
executing a specific script with pre-defined parameters. 
Usually, scripts are internal automatic investigation tools for a service, and only the service team has access to the tools.
 
For instance, in Figure~\ref{fig:model}, the ``Known issue?'' action node queries the database to see whether the current incident is a known one or not based on its alert messages.
If it is a known issue, execution flow will enter the ``True'' branch to give mitigation actions directly.
Otherwise, a query script that can aggregate threads with the same stack traces will be executed.     
It will obtain an instantaneous list of the stacks on 
all the managed threads in the target process and 
then group common stacks together in order to 
identify potential deadlocks/blocking code paths in the process.
 
The query action can also output an enum value to 
decide the next action node to execute, e.g., after getting the top error message on the exception stack traces, i.e., "Get top error msg" node, 
the next action node to be run depends on the exception type. 
Based on the error messages, a specific team will be reported 
and engaged, as shown in Figure~\ref{fig:model}.

\textbf{Mitigation action}: This action refers to the strategic steps suggested to alleviate an incident, such as ``restart service'' or ``engage other teams'', as depicted in Figure~\ref{fig:model}.
It's important to note that handlers do not always provide exact mitigation strategies for every incident, due to handlers' pre-defined nature, which may not cover all possible situations. For instance, Incident 4 in Table~\ref{tbl:cloud_incidents}, categorized under code regression, presents a case where identification and rectification of such code issues can be challenging. In cases where the incident handler is uncertain, it will offer intermediate diagnostic information to the OCEs without mitigation.
 
\subsubsection{Multi-source diagnostic information} \name{}'s diagnostic information collection stage serves as a valuable tool for OCEs by aggregating data from a myriad of sources. OCEs only need to customize the action in the handler to acquire the diagnostic information from a target source. For instance, as illustrated in Figure~\ref{fig:probe_error_log}, \name{} can assimilate diverse data such as error logs, exception stack traces, and socket metrics related to a specific incident.
The error log and exception stack trace alone does not provide sufficient insight to identify the root cause of the incident.
However, when supplemented with the socket metrics, a more comprehensive picture emerges. In this example, it is clear that the UDP socket is exhausted, which is the root cause.
 
In the case of new incidents, \name{} can perform a range of common checks, such as evaluating the provisioning status or analyzing thread stacks. This assists OCEs in gaining a holistic understanding of the situation. Note that the information collected is pre-defined in the actions of the \name{} handler, ensuring that only relevant data is gathered, thus avoiding  overwhelming information that is unnecessary. By providing this comprehensive diagnostic information, \name{} empowers OCE teams to troubleshoot issues efficiently. 
They can use the gathered information as guidance to address incidents more effectively.

\begin{figure}[h]
\centering
\begin{tcolorbox}[colback=white, colframe=black, width=\columnwidth, boxrule=1pt, sharp corners]
\footnotesize
\ttfamily
DatacenterHubOutboundProxyProbe \textbf{probe log result} from [MachineID]. \\
Total  Probes: 2, Failed Probes: 2\\
\noindent\begin{tabular}{lllll}
Id & Level & Created & & Description \\
-- & ----- & ------- & & ----------- \\
2 & Error & 11/21/2022 & 2:04:20 AM & Probe result \\
2 & Error & 11/21/2022 & 1:49:20 AM & Probe result \\
\end{tabular}
 
\noindent Failed probe error:\\
Name: No such host is known. \\
A WinSock error: 11001 encountered when connecting to host: [HOST NAME] \\
Count: 2\\
\dots{}\\
\noindent \textbf{Exceptions}:\\
InformativeSocketException: No such host is known. \\
A WinSock error: 11001 encountered when connecting to host: [HOST NAME] \\
\quad at TcpClientFactory.Create(...)\\   
\quad at SimpleSmtpClient.Connect(...)\\
\dots{}\\
\noindent \textbf{Total UDP socket count}: \textbf{\underline{15276}}\\
Total UDP socket count by process and processId (top 5 only):\\
14923: \service{}.exe, 203736 \\                
15: w3wp.exe, 102296 \\                         
8: svchost.exe, 4748  \\                   
7: \company{}.\service{}.Store.Worker.exe, 74060 \\
7: \company{}.\service{}.Store.Worker.exe, 87724 
\end{tcolorbox}
\caption{Diagnostic information for hub port exhaustion. 
}
\label{fig:probe_error_log}
\end{figure}

\subsection{LLMs for Incident Explanation}
\label{llm}

Upon thorough investigation, each incident within our service 
is manually assigned a root cause category by our seasoned OCEs.
OCEs will use the categories to classify the historical incidents
and guide the new incoming incidents' RCA.
However, reasoning the incidents and inferring their categories are time-consuming and potentially overwhelming for OCEs, who have a tight time budget.
Given this, we have identified the categorization of incident root causes as our primary downstream task.
 
Recently, LLMs have demonstrated remarkable capabilities in understanding the context 
of downstream tasks and generating relevant information from demonstrations, 
making them a possible choice for incident RCA.
However, reasoning the incident root cause is not a simple task, 
and LLMs may not be able to achieve the optimal results on long-tail 
or domain-specific tasks without any guidance~\cite{chalkidis2023chatgpt, kasai2023evaluating}.
Chain-of-Thoughts (CoT) prompting is a gradient-free technique that elicits 
LLMs to generate intermediate reasoning steps that lead to the final answer.
In few-shots CoT prompting, a few manual demonstrations 
that are composed of a question and a reasoning chain 
that leads to an answer for each of them.
Inspired by the above ideas, diagnostic information provided by \name{} handlers can be used as ingredients 
for the reasoning process of the incidents.

\subsubsection{Embedding model.}
\label{sec:embedding}
Our observation is that the \textit{semantics of incidents can be revealed from the context in which the diagnostic information is described}. A common approach to extracting such contextual semantics involves the use of embedding models. The objective is to map the diagnostic information into an embedding space (i.e., numeric vector space), where the distances between vectors represent the semantic similarity of incidents. Choosing a computationally efficient embedding model allows us to preserve accuracy while handling a large number of incidents.
 
We employ FastText as our embedding model, which is efficient, insensitive to text input length, and generates dense matrices, making it easy to calculate the Euclidean distance between similar vectors. Furthermore, since our downstream task is domain-specific to the incident root cause reasoning, and the incident-related information is internal to our company, we opt to train a FastText model on our historical incidents rather than using a pre-trained large language model as our embedding model, which is costly and inefficient. Additionally, we provide users with the flexibility to customize their embedding model if desired.
 
\subsubsection{Nearest neighbor search.}
Incidents are heterogeneous, making it impractical to combine all past incidents' information for sampling due to the prompt length limitations, even after summarization.
To selectively choose past cases as samples in the prompt,
we design a new similarity formula: 
$$
Distance(a,b)= ||a-b||_2 \\
$$
$$
Similarity(a,b)=\frac 1 {1+Distance(a,b)}*e^{-\alpha{|T(a)-T(b)|}}
$$
to calculate the similarity between two incidents.
It first computes the Euclidean distance for every pair of incident vectors. 
Importantly, it also takes into account the temporal distance between incidents, reflecting our Insight-2 in Section~\ref{insight:2}.
Here, $T(x)$ stands for the date of incident $x$.
This consideration of temporal distance is crucial as it influences the relevance of past incidents to the current ones.
After calculating similarities, we select the top $K$ incidents from different categories as demonstrations for the LLM. This approach ensures a diverse and representative set of incidents for effective LLM reasoning.
The values of $\alpha$ and $K$ have been determined as 0.3 and 5, respectively, through empirical evaluation, as will be presented in Section~\ref{sec:comparison}.

\subsubsection{Diagnostic information summary.}
LLMs have shown potential for automatic summarization~\cite{mastropaolo2021studying}.
Nonetheless, the length of the diagnostic information collected from \name{} handlers
is often too extensive. 
As shown in Figure~\ref{fig:probe_error_log}, the diagnostic information of an incident can have more than 2000 tokens with low readability of the log messages.
The considerable number of tokens in the incident description can 
pose challenges for the LLM to effectively process 
and may introduce noise. 
Therefore, feeding the diagnostic information of an incident directly into the LLM 
to make a prediction could not be an ideal choice, 
let alone using the information from multiple sources.
In this regard, we add another layer to 
leverage the LLM's ability to summarization 
to summarize the diagnostic information first
before making the diagnosis reasoning.
We construct the prompt in the way of Figure~\ref{fig:prompt_summary}. We ask LLM to summarize the diagnostic information into 120-140 words without outputting any unrelated information.
This summarization process makes the diagnostic information more concise and informative, 
which forms the basis for the later CoT prompting.
Figure~\ref{fig:log_summary} illustrates a more readable and concise text generated by \name{}, which is a summary (113 tokens) of the previous diagnostic information example in Figure~\ref{fig:probe_error_log}, highlighting the key details such as the number of UDP ports used and the process utilizing the most. 
Specifically, we employ the tiktoken~\cite{tiktoken} tokenizer to count text tokens.

\begin{figure}[h]
\centering
\fbox{
\begin{minipage}{.9\columnwidth}
\textit{``Please summarize the above input. Please note that the above input is incident diagnostic information. The summary results should be about 120 words, no more than 140 words, and should cover important information as much as possible. Just return the summary without any additional output.''}
\end{minipage}
}
\caption{Prompt to summarize diagnostic information.}
\label{fig:prompt_summary}
\end{figure}

\begin{figure}[h]
\centering
\fbox{
\begin{minipage}{.9\columnwidth}
\textit{``The DatacenterHubOutboundProxyProbe has failed twice on the backend machine, with both failures due to a WinSock error 11001 indicating that the host is unknown. This error was encountered while attempting to connect to the host. The error is associated with the EOP service and has not been notified yet. The failure context suggests the same issue. \textbf{The total UDP socket count is \underline{15276}, with the majority being used by the \underline{\service{}.exe} process.} The issue seems to be related to the SMTP connection and requires further investigation.''}
\end{minipage}
}
\caption{The summarized diagnostic information.}
\label{fig:log_summary}
\end{figure}

\subsubsection{Prediction prompt construction.}
CoT prompting is a gradient-free technique that guides LLMs to produce intermediate reasoning steps leading to the final answer.
In few-shot CoT prompting, several demonstrations include a question and a reasoning chain that directs the answer.
By drawing inspiration from automatically constructing the prompt to form the reasoning chains~\cite{zhang2023automatic}, we can view the summarized diagnostic information and the labeled root cause categories as questions and reasoning,
so finding the nearest incident neighbor is the automatic reasoning chain construction, 
aligning with the CoT prompting context well. 
Note that we use the original incident information to do the embedding and nearest neighbor search, 
and use the corresponding summarized information as part of demonstrations in the prompt.
We construct the prompt like Figure~\ref{fig:incident_selection_prompt} to ask the LLM to choose the most likely incident that has the same root cause as the current incident, and also we explicitly push the LLM to reason by using ``give your explanation'' indications in the prompt.

\begin{figure}[h]
\centering
\fbox{
\begin{minipage}{.9\columnwidth}
\textbf{Context:} The following description shows the error log information of an incident. Please select the incident information that is most likely to have the same root cause and \textbf{give your explanation} (just give one answer). If not, please select the first item ``Unseen incident''.\\
\textbf{Input:} The DatacenterHubOutboundProxyProbe probe result from [BackEndMachine] is a failure ...\\
Options:\\
\hspace*{0.4cm}\textbf{A:} Unseen incident.\\
\hspace*{0.4cm}\textbf{B:} The DatacenterHubOutboundProxyProbe has failed twice ... \textit{category: \textbf{HubPortExhaustion}}.\\
\hspace*{0.4cm}\textbf{C:} There are 62 managed threads in process \service{}Delivery ...
\textit{category: \textbf{AuthCertIssue}}.
\end{minipage}
}
\caption{The prompt to predict incident category.}
\label{fig:incident_selection_prompt}
\end{figure}

\subsection{Implementation}
 
We have developed and deployed \name{} using a combined total of 58,286 lines of code, consisting of 56,129 lines of C\# and 2,157 lines of Python.
 
\begin{figure}[h]
\centering
\includegraphics[width=0.4 \textwidth]{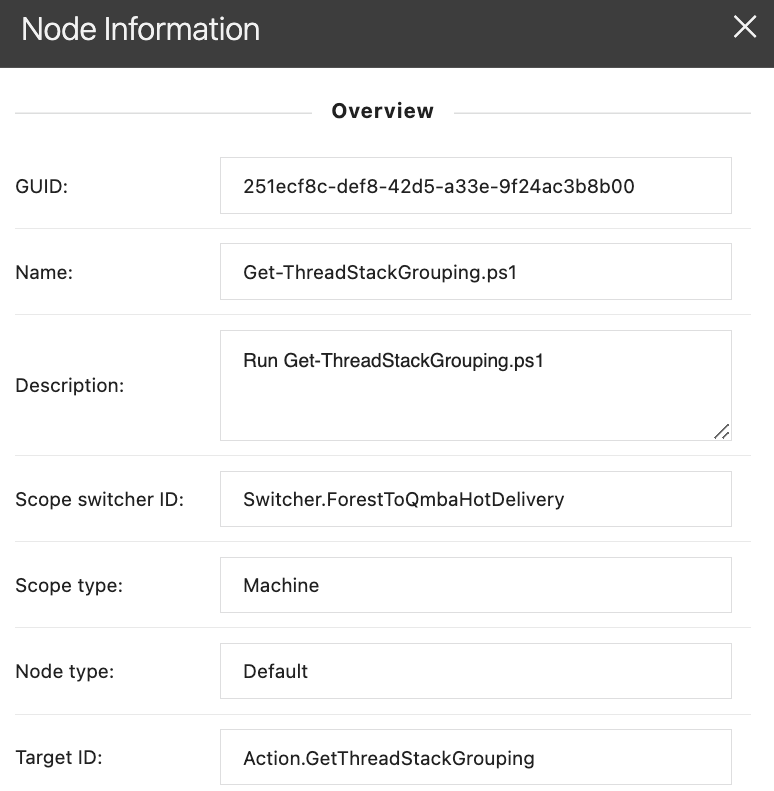}
\caption{
Web-based user interface of \name{} for handler construction. 
}
\label{fig:gui_eval}
\end{figure}

To facilitate the building of the \name{} incident handler, 
we have implemented \name{}'s handler construction as a web application
as shown in Figure~\ref{fig:gui_eval}.
To support a new type of alert in \name{}, OCEs only need to
add a new handler in the handler construction GUI
according to her expertise. 
After the new handler has been constructed,
it will be stored in the database, and
OCEs can modify it by creating new action nodes 
or deleting old nodes.

\section{Evaluation}
\label{sec:evaluation}

We aim to answer the following questions in our evaluation:
\begin{enumerate}[label=(\arabic*), leftmargin=*]
\item How effective and efficient is \name{} as an on-call system when predicting root cause categories and assisting OCEs? \name{} achieves 0.766 and 0.533 for Micro-F1 and Macro-F1 separately when predicting the root cause category of cloud incidents, outperforming all our baselines with a low running overhead (4.205 seconds). \name{} is also able to generate new root cause category labels for unseen incidents with explanations.
 
\item How do different components of \name{} facilitate its diagnosis and prediction? \name{} has proven that the diagnostic information collection component, GPT summarization, and chain-of-thoughts prompting all contribute to \name{}'s prediction effectiveness. 
 
\item Is \name{} suitable for deployment in real production services, and are \name{}'s results trustworthy? \name{}'s diagnostic information collection module has been deployed across 30 teams within \company{} for over four years. To evaluate the trustworthiness of \name{}, each experiment was conducted over three rounds, and \name{} can consistently achieve a high Micro-F1 score of over 0.70 and a Macro-F1 score exceeding 0.50.
\end{enumerate}

All experiments are performed on the server with Intel(R) Core(TM) i7-9700 CPU @ 3.00GHz, 32.0 GB physical memory, and Intel UHD Graphics 630. The OS of the server is Windows 11 Enterprise.

\subsection{Target System and Dataset}
We evaluate \name{} in a global email service system named \service{} within the \company{}. 
The \service{} team focuses on developing and maintaining the components responsible for mail flow, routing, and delivery. This system interacts with various other services to ensure seamless integration with a multitude of products and services, including serviceA, serviceB, and serviceC. Hence, it is representative of complex, real-world systems that interact with multiple components.
With around 150 billion messages being delivered daily, \service{} operates at a colossal scale and caters to customers worldwide, adding another layer of diversity and complexity.
The system ensures the secure and effective transmission of emails between users, utilizing various protocols such as SMTP, IMAP, and POP3. Given its crucial role in communications infrastructure, it is essential to have effective and efficient incident management capabilities.
 
We collect a one-year dataset of 653 incidents from \company{}'s \service{} service 
to investigate \name{}'s efficacy in practice. 
It is important to note that each of these incidents represents complex issues in a large-scale, globally distributed system, and thus each provides valuable insights.
The dataset is manually labeled with root cause categories by experienced OCEs, which serves as our ground truth. 
We divide the incidents into train (75\%) and testing sets (25\%).
 
We conduct experiments on two large language models in \name{}, \ie GPT-3.5-turbo, and GPT-4 (8K tokens), which are the latest models from OpenAI. 
We choose GPT-4 as the default model in \name{} because it has the best performance.
 
\subsection{Compared Approaches}
We have selected XGBoost, FastText, and fine-tuned LLMs as our baselines to compare with \name{}. After training or fine-tuning with the training dataset, we directly apply these approaches to the testing set to do the classification task. We have also made another two variants, i.e., GPT-4 Prompt and Embed. to evaluate the design of \name{}.  
 
\begin{itemize}[leftmargin=*]
\item \textbf{XGBoost} provides a parallel tree boosting that has been commonly used in the networking system diagnosis.
 
\item \textbf{FastText} is a popular lightweight textual embedding approach, which has been adopted in testbed studies with fault injections for root cause diagnosis tasks. We directly apply FastText to our dataset to do the classification.    
 
\item \textbf{Fine-tune GPT} is to fine-tune a pre-trained GPT-3.5 model with our training dataset and evaluate its performance on our testing dataset with the temperature parameter set to 0. It does not use a prompt design (i.e., CoT prompting) like \name{} but directly predicts the category with the original diagnosis information. Note that GPT-4 is currently not available for fine-tuning. 
 
\item \textbf{GPT-4 Prompt} is a variant of \name that directly predict category with \name{}'s diagnosis information summaries. Its prompt only contains the incident being predicted, so there is no historical incident information as demonstrations.
 
\item \textbf{GPT-4 Embed.} is a variant of \name that changes the embedding model from FastText to GPT embedding.

\end{itemize}
 
\begin{table}[ht]
\caption{Effectiveness of different methods.}
\vspace{-7.5pt}
\centering
\small
\begin{tabular}{@{}lS[table-format=1.3]S[table-format=1.3]cS[table-format=2.3]@{}}
\toprule
\textbf{Method} & \multicolumn{2}{c}{\textbf{F1-score}} & \multicolumn{2}{c}{\textbf{Avg. Time (s)}}\\
\cmidrule(lr){2-3} \cmidrule(l){4-5}
&\textbf{Micro} & \textbf{Macro} & \textbf{Train.} & \textbf{Infer.}  \\
\midrule
FastText \cite{zhao2023robust} & 0.076 & 0.004 & 10.592 & 0.524 \\
XGBoost \cite{arzani2016taking} & 0.022  & 0.009 & 11.581 & 1.211  \\
Fine-tune GPT \cite{ahmed2023recommending} & 0.103 & 0.144 & 3192 & 4.262 \\ \midrule
GPT-4 Prompt & 0.026 & 0.004 & {--} & 3.251\\
GPT-4 Embed. & 0.257 & 0.122 & 1925 & 3.522 \\ \midrule
\name{} (GPT-3.5) & 0.761 & 0.505 & 10.562 & 4.221\\
\textbf{\name{} (GPT-4)} & \textbf{0.766} & \textbf{0.533} & 10.562 & 4.205\\
\bottomrule
\end{tabular}
\label{tab:combined_comp}
\end{table}

\subsection{Effectiveness and Efficiency}
\label{sec:effectiveness}
We evaluate \name{}'s effectiveness by predicting the root cause category of an incident based on the summarized diagnostic information using micro and macro F1-score metrics. These metrics calculate the harmonic mean of the precision and recall. The micro F1-score aggregates the performance of all classes, taking into account the contribution of each sample, while the macro F1-score focuses on the performance of each individual class. \name{} achieves a micro F1-score of 0.766 and a macro F1-score of 0.533 on our testing dataset.

As shown in Table~\ref{tab:combined_comp}, \name{} outperforms other approaches, and it tends to incur an acceptable higher runtime overhead. 
The performance of baseline approaches is poor, since multiple root cause categories exhibit a long tail (imbalanced) distribution, as shown in Figure~\ref{fig:incident_frequency}, and traditional machine learning models (FastText and XGBoost) and fine-tuning GPT model need a large amount of training data to produce accurate predictions.
Directly employing GPT-4 prompt or GPT-4 embedding approach without our design lacks domain-specific knowledge for GPT-4 to make decisions. 
On the contrary, \name{} leverages the powerful LLM to learn the domain-specific knowledge from minimal cases, so that it can achieve the best performance. 
Results indicate that \name{} not only provides higher accuracy but also maintains a reasonable level of efficiency, making it a suitable choice for incident root cause analysis. 
 
When facing incidents that \name{} has never seen before, \name{} is capable of generating a new category keyword to depict the new incident case. For example, Incident 8 in Table~\ref{tbl:cloud_incidents} is a new incident case that \name{} has never encountered. \name{}'s prediction component is able to predict it as a new category ``I/O Bottleneck''. Although OCEs subsequently categorize it as ``DiskFull'' in post-investigation, the fundamental aspects of the problem identified by \name{} align closely with the human-derived label. The corresponding \name{}'s explanation, illustrating how it arrived at the "I/O Bottleneck" categorization, is provided in Figure~\ref{fig:explanation}.

\begin{figure}[h]
\centering
\fbox{
\begin{minipage}{.9\columnwidth}
\textit{The prediction of ``I/O Bottleneck'' was made based on the occurrence of System.IO.IOExceptions within crucial functions handling input/output operations, suggesting an issue with data processing. The nested exception within the DiagnosticsLog module reinforces this notion. These errors, combined with crashes on different backend machines, point to a system struggle with handling data flow.}
\end{minipage}
}
\caption{\name{}'s explanation of an incident.}
\label{fig:explanation}
\end{figure}
 
\begin{figure}
\centering
\begin{subfigure}[h]{0.5\columnwidth} 
\centering
\includegraphics[width=\textwidth]{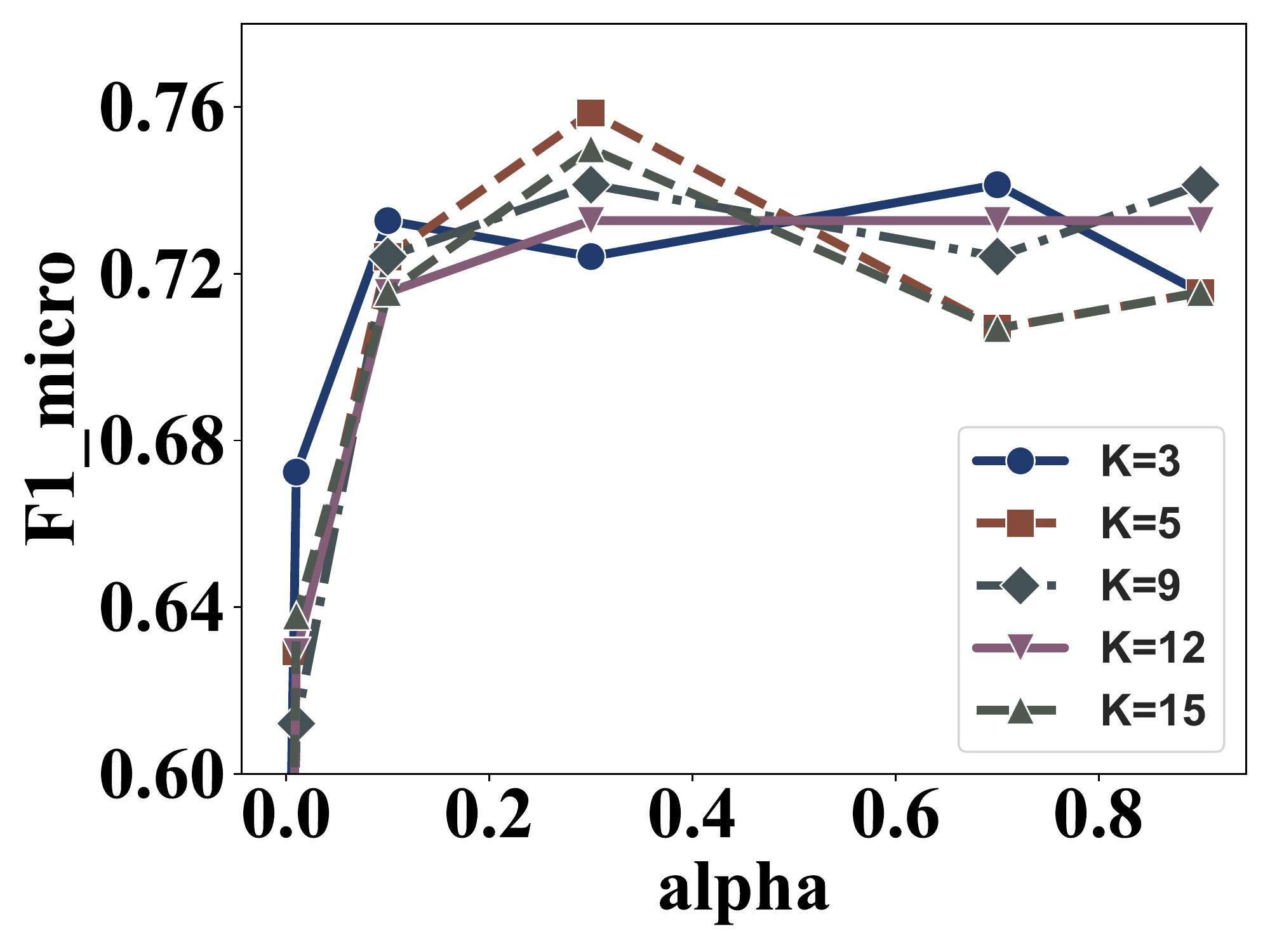}
\caption{F1 micro.}
\label{fig:F1_micro}
\end{subfigure}
~
\begin{subfigure}[h]{0.5\columnwidth}
\centering
\includegraphics[width=\textwidth]{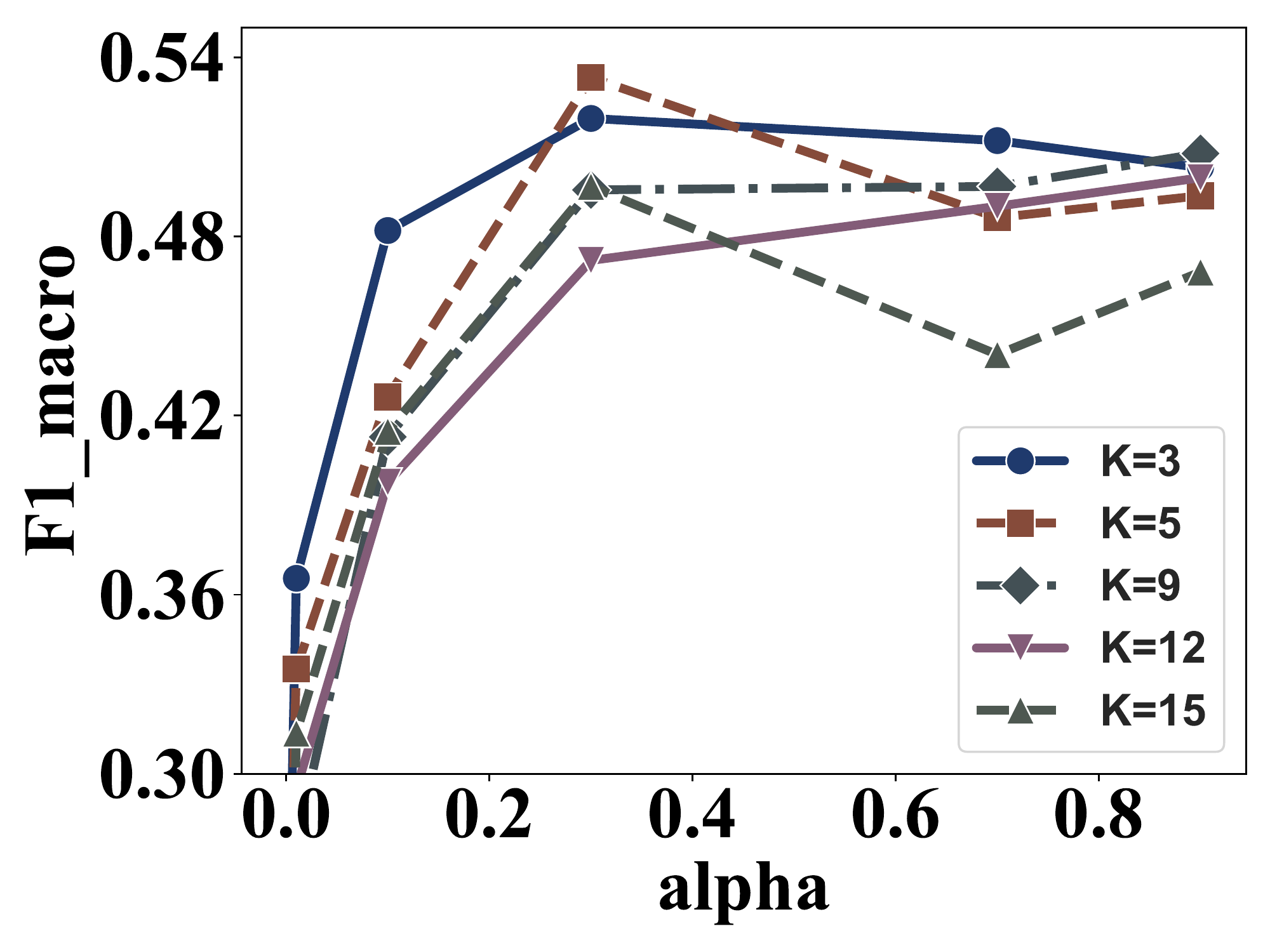}
\caption{F1 macro.}
\label{fig:F1_macro}
\end{subfigure}
\caption{Effectiveness of using different K and alpha. 
} 
\label{fig:effectiveness} 
\end{figure}

\subsection{Comparison Analysis}
\label{sec:comparison}
To understand how different components of \name{} facilitate root cause analysis, we conduct an ablation study on the different \name{}'s components.

\textbf{Evaluation on diagnostic information.}
First, we evaluate the impact of diagnostic information on effectiveness. In particular, we compare diagnostic information collected from the collection stage with other different incident-related information, namely, incident alert information and \name{} handler action output.
AlertInfo includes the alert type and alert scope. Alert type is a pre-defined anomaly description from a monitor, which only reflects a symptom of the incident instead of the root cause, e.g., an exception type from external monitors. The alert scope is the scope of the incident, e.g., a single machine.
ActionOutput is the output of a series of executed \name{} actions, which are hashed as key-value pairs.
As shown in Table~\ref{tab:sum_comp}, using diagnostic information alone can outperform others in both Micro-F1 (0.689) and Macro-F1 scores (0.510). 
The interesting observation here is that mixing the diagnostic information with others will not enhance \name{}'s predictive capabilities.
This demonstrates that an excess of information can negatively impact the LLM's prediction performance.

\begin{table}
\caption{Effectiveness of different prompt context for \name{}. \ding{51}\textsuperscript{\footnotesize{sum.}} stands for the summarized diagnostic information.
}
\vspace{-0.75pt}
\centering
\small
\begin{tabular}{cccccc}    
\toprule
\multicolumn{3}{c}{\textbf{Data Source}} & \multicolumn{2}{c}{\textbf{F1-score}}\\
\cmidrule[0.5pt](lr){1-3} \cmidrule[0.5pt](lr){4-5}
AlertInfo & DiagnosticInfo & ActionOutput & Micro & Macro\\
\midrule
& \ding{51} & & 0.689 & 0.510 \\
& \ding{51}\textsuperscript{\footnotesize{sum.}} & & \textbf{0.766} & \textbf{0.533} \\ \hline
\ding{51} & & & 0.379 & 0.245\\
\ding{51} & \ding{51} & & 0.525 & 0.511\\
\ding{51} & & \ding{51} & 0.431 & 0.247\\
& \ding{51} & \ding{51} & 0.501 & 0.449 \\
\ding{51} & \ding{51} & \ding{51} & 0.440 & 0.349\\
\bottomrule
\end{tabular}
\label{tab:sum_comp}
\vspace{-7.5pt}
\end{table}
 
\textbf{Evaluation on GPT summarization.}
We evaluate the role of GPT summarization in enhancing \name{}'s effectiveness. 
As depicted in Table~\ref{tab:sum_comp}, utilizing summarized diagnostic information leads to the highest Micro-F1 and Macro-F1 scores, marking improvements of 0.077 and 0.023, respectively, over the non-summarized diagnostic information.
The results demonstrate that the summarization step effectively condenses the information, allowing for more efficient and accurate processing of incident data.
 
\textbf{Evaluation on few-shots CoT reasoning.}
We assess how few-shots CoT reasoning contributes to improving effectiveness. GPT-4 Prompt approach in Table~\ref{tab:combined_comp}, which directly predicts the category without any sample, only achieves 0.026 and 0.004 for Micro-F1 and Macro-F1 respectively.
As shown in Figure~\ref{fig:F1_micro} and Figure~\ref{fig:F1_macro}, we compare the performance of \name{} with different numbers of samples in the Chain-of-thoughts reasoning.
Our analysis reveals that the best combination of the number of samples and alpha values are 5 and 0.3, which achieves the highest F1 scores.
Note that more samples in the CoT reasoning do not always incur an improvement for \name{},
and the value of the alpha plays an important role in deciding the effectiveness. 
When the alpha is appropriate, 
it allows \name{} to better capture the time relationships between different incidents, leading to more accurate predictions.

\subsection{Deployment Status and Scale}
 
We have successfully deployed \name{}'s diagnostic information collection module across over 30 teams within \company{}, where it has been in active use for over four years. The system is tailored to each team's specific requirements, with custom handlers built for each unique setting. Not all handlers are currently enabled in the production environment, as some are still under development and rigorous testing. 
We select the top 10 teams that utilize the most \name{} incident handlers as shown in Table~\ref{table:team_summary_scale}.
We observe that the average running time for each incident ranges from 15 seconds to 841 seconds. 
The highest running time is attributable to the team's large-scale and complex system infrastructure.
The root cause prediction module has also been rolled out in the \service{} service.  
 
As part of our commitment to continuous improvement and quality user experience, we have incorporated a feedback mechanism in incident notification emails to get user perspectives from OCEs.
According to the collected feedback, 
most OCEs expressed satisfaction with \name{}.
Despite the manual effort involved in creating a new incident handler,
OCEs find the process convenient when reusing and modifying handler actions from the database.
\name{} is able to save OCEs a significant amount of time to collect diagnostic information, triage incident, perform mitigation and do postmortem analysis.

\begin{table}[htbp]
\caption{Teams in \company{} using \name{} to automatically collect diagnostic information.}
\vspace{-7.5pt}
\centering
\begin{tabular}{lcc}
\toprule
\multirow{2}{*}{Team} & Avg. exec. & \# Enabled  \\
& time (seconds) & handler  \\ 
\midrule
Team 1 & 841 & 213 \\
Team 2 & 378 & 204 \\
Team 3 & 106 & 88 \\
Team 4 & 449 & 42 \\
Team 5 & 136 & 41 \\
Team 6 & 91 & 34 \\
Team 7 & 449 & 32 \\
Team 8 & 255 & 32 \\
Team 9 & 323 & 31 \\
Team 10 & 22 & 18 \\
\bottomrule
\end{tabular}
\label{table:team_summary_scale}
\end{table}

\subsection{Tustworthiness}
\label{sec:trustworthiness}
While GPT has shown great potential and impressive results in various tasks, it is known to exhibit some instability in certain complex tasks such as question answering, as noted by Tan et al.~\cite{tan2023evaluation}. These instabilities could potentially lead to variable results. In order to ensure the trustworthiness and stability of the GPT's predictive capabilities in \name{}, each experiment has been conducted three rounds. In each round, \name{} was able to maintain a high level of performance, with the Micro-F1 consistently above 0.70 and the Macro-F1 remaining above 0.50.

\section{Discussion}
\label{sec:discussions}

\name{}'s effectiveness depends on the ability of the LLM. Currently, \name{} is only integrated with OpenAI's GPT models, and we have not yet explored the potential effectiveness of other available LLMs. As such, the model's performance may vary depending on the strengths and weaknesses of the specific LLM employed. 
 
We conducted our evaluation of \name{}'s prediction module using the incident dataset from \service{}. The dataset was prepared with the assistance of experts in \team{}, given their extensive experience and established practice of incident labeling. Note that
the effectiveness of \name{} is also influenced by the quality of the root cause category labels written by human. Currently, all root cause categories are manually labeled by our experienced OCEs.
\name{}'s diagnosis information collection has been deployed in over 30 teams. Consequently, a valuable future work would be to evaluate \name{} across different services to gain a more comprehensive understanding of its generalizability and adaptability.

The handler in \name{} is designed to initiate responses based on alerts from monitors/watchdogs. This ensures that when there is a designated incident handler for a particular alert type, it gets activated with an accuracy rate of 100\%. Nevertheless, it's crucial to highlight that \name{}'s capabilities are constrained in scenarios where the monitors fail to detect an incident, or when there is an absence of a corresponding handler for a particular incident. This, in turn, limits the applicability of \name{}.

We conducted three rounds of experiments to evaluate \name{}'s effectiveness. However, the occasional instability of LLMs can influence their effectiveness, causing variations across different rounds. Another potential threat to internal validity lies in the implementation of our approach and those we compared against. To mitigate this risk, two authors have carefully checked the code. In particular, our implementation is based on the matured frameworks.

\section{Related Work}
\label{sec:related}
 
\point{Root cause analysis}
Root cause analysis in large cloud services has become a popular topic of research in the system and software engineering communities \cite{zhang2021understanding, alquraan2018analysis, gao2018empirical, ghosh2022fight, liu2019bugs, yuan2014simple, ma2020diagnosing, chen2019understanding, leesatapornwongsa2017scalability, lou2020understanding}.
It aims to identify the root causes of failures and performance issues based on various data sources, such as metrics, logs, and traces.
Previous studies have proposed different approaches for root cause analysis using one of these data sources. 
For example, some methods rely on metrics to  extract failure patterns \cite{ma2020diagnosing,  zhang2021cloudrca} or to construct service dependency graphs \cite{ma2020automap, li2022mining}.
Others use logs to analyze a subset of log messages \cite{zhang2021onion, ahmed2023recommending} or to examine the details within each log message \cite{zhang2023system, li2023did}.  
Moreover, some techniques utilize trace to locate the faulty service \cite{xie2023unsupervised, TraceArk, li2021practical, liu2021microhecl}. 
Different from prior work, we build a system that can automatically integrate metrics, logs, and traces for root cause analysis with state-of-the-art large language models.

\point{Large Language Models}
In recent years, the rise of LLM has brought new opportunities to the field of software systems by enabling various tasks such as code generation, summarization, repair, testing, and root cause analysis   \cite{mastropaolo2021studying, mastropaolo2022using, fu2022vulrepair,ahmed2023recommending}.
For example, Mastropaolo \etal \cite{mastropaolo2021studying} studied the ability of fine-tuned T5 in the following tasks: automatic bug fixing, generation of assert statements, code summarization, and injection of code mutants.
LANCE \cite{mastropaolo2022using} uses fine-tuned T5 to automatically generate logging statements for Java methods.
VulRepair \cite{fu2022vulrepair} also fine-tune T5 on vulnerability repairs datasets to automatically propose vulnerability fixes.
Zhang \etal \cite{zhang2022using} proposes to use prompting for LLM to improve code version control.
Ahmed \etal \cite{ahmed2023recommending} fine-tune GPT-x models to recommend root causes and mitigation steps to facilitate cloud incident management.
In contrast to previous studies, \name employs advanced LLMs to summarize diagnosis data and leverage the chain-of-thoughts ability to predict and explain root causes.

\section{Conclusion}
\label{sec:conclusion}

\name{} represents a pioneering tool in the realm of cloud incident management, facilitating efficient root cause analysis for OCEs. 
It introduces a unique approach to multi-source data collection through its diagnostic information collection stage, utilizing predefined incident handlers. These handlers, constructed by OCEs, systematically gather multi-source diagnostic information, which sets the foundation for the subsequent analysis.
Furthermore, \name{} integrates a large language model in its root cause prediction stage. This model autonomously processes the collected diagnostic data, predicting and explaining the root cause category. This integration of AI techniques into cloud incident management demonstrates the potential of \name{} in enhancing the efficiency and accuracy of root cause analysis.
 
\section*{Acknowledgement}
 
We thank our shepherd, Ang Chen, and the anonymous reviewers for their insightful comments.
We thank Ning Ding, Xupei Wang, and Zhaoying Li for their participation, support and contributions to the \name{} project.
We thank all the on-call engineers within Microsoft who engaged with us.

\bibliographystyle{acm}
\balance
\bibliography{references}

\begin{thebibliography}{10}

\bibitem{ahmed2023recommending}
{\sc Ahmed, T., Ghosh, S., Bansal, C., Zimmermann, T., Zhang, X., and Rajmohan,
  S.}
\newblock Recommending root-cause and mitigation steps for cloud incidents
  using large language models.
\newblock In {\em Proceedings of the 45th International Conference on Software
  Engineering (ICSE'23)\/} (2023).

\bibitem{alquraan2018analysis}
{\sc Alquraan, A., Takruri, H., Alfatafta, M., and Al-Kiswany, S.}
\newblock An analysis of network-partitioning failures in cloud systems.
\newblock In {\em Proceedings of the 13th USENIX Conference on Operating
  Systems Design and Implementation (OSDI'18)\/} (2018).

\bibitem{arzani2016taking}
{\sc Arzani, B., Ciraci, S., Loo, B.~T., Schuster, A., and Outhred, G.}
\newblock Taking the blame game out of data centers operations with netpoirot.
\newblock In {\em Proceedings of the 2016 ACM SIGCOMM Conference
  (SIGCOMM'16)\/} (2016).

\bibitem{bansal2020decaf}
{\sc Bansal, C., Renganathan, S., Asudani, A., Midy, O., and Janakiraman, M.}
\newblock Decaf: Diagnosing and triaging performance issues in large-scale
  cloud services.
\newblock In {\em Proceedings of the ACM/IEEE 42nd International Conference on
  Software Engineering: Software Engineering in Practice\/} (2020).

\bibitem{brown2020language}
{\sc Brown, T., Mann, B., Ryder, N., Subbiah, M., Kaplan, J.~D., Dhariwal, P.,
  Neelakantan, A., Shyam, P., Sastry, G., Askell, A., et~al.}
\newblock Language models are few-shot learners.
\newblock {\em Advances in neural information processing systems\/} (2020).

\bibitem{chalkidis2023chatgpt}
{\sc Chalkidis, I.}
\newblock Chatgpt may pass the bar exam soon, but has a long way to go for the
  lexglue benchmark.
\newblock {\em arXiv preprint arXiv:2304.12202\/} (2023).

\bibitem{chen2019understanding}
{\sc Chen, H., Dou, W., Jiang, Y., and Qin, F.}
\newblock Understanding exception-related bugs in large-scale cloud systems.
\newblock In {\em 2019 34th IEEE/ACM International Conference on Automated
  Software Engineering (ASE'19)\/} (2019).

\bibitem{chen2019empirical}
{\sc Chen, J., He, X., Lin, Q., Xu, Y., Zhang, H., Hao, D., Gao, F., Xu, Z.,
  Dang, Y., and Zhang, D.}
\newblock An empirical investigation of incident triage for online service
  systems.
\newblock In {\em 2019 IEEE/ACM 41st International Conference on Software
  Engineering: Software Engineering in Practice (ICSE-SEIP'19)\/} (2019).

\bibitem{chen2019continuous}
{\sc Chen, J., He, X., Lin, Q., Zhang, H., Hao, D., Gao, F., Xu, Z., Dang, Y.,
  and Zhang, D.}
\newblock Continuous incident triage for large-scale online service systems.
\newblock In {\em 2019 34th IEEE/ACM International Conference on Automated
  Software Engineering (ASE'19)\/} (2019).

\bibitem{chen2020incidental}
{\sc Chen, J., Zhang, S., He, X., Lin, Q., Zhang, H., Hao, D., Kang, Y., Gao,
  F., Xu, Z., Dang, Y., et~al.}
\newblock How incidental are the incidents? characterizing and prioritizing
  incidents for large-scale online service systems.
\newblock In {\em Proceedings of the 35th IEEE/ACM International Conference on
  Automated Software Engineering (ASE'20)\/} (2020).

\bibitem{chen2021evaluating}
{\sc Chen, M., Tworek, J., Jun, H., Yuan, Q., Pinto, H. P. d.~O., Kaplan, J.,
  Edwards, H., Burda, Y., Joseph, N., Brockman, G., et~al.}
\newblock Evaluating large language models trained on code.
\newblock {\em arXiv preprint arXiv:2107.03374\/} (2021).

\bibitem{chen2023rainmaker}
{\sc Chen, Y., Sun, X., Nath, S., Yang, Z., and Xu, T.}
\newblock {Push-Button Reliability Testing for Cloud-Backed Applications with
  Rainmaker}.
\newblock In {\em Proceedings of the 20th USENIX Symposium on Networked Systems
  Design and Implementation (NSDI'23)\/} (2023).

\bibitem{fu2022vulrepair}
{\sc Fu, M., Tantithamthavorn, C., Le, T., Nguyen, V., and Phung, D.}
\newblock Vulrepair: a t5-based automated software vulnerability repair.
\newblock In {\em Proceedings of the 30th ACM Joint European Software
  Engineering Conference and Symposium on the Foundations of Software
  Engineering (ESEC/FSE'22)\/} (2022).

\bibitem{ganatra2023detection}
{\sc Ganatra, V., Parayil, A., Ghosh, S., Kang, Y., Ma, M., Bansal, C., Nath,
  S., and Mace, J.}
\newblock Detection is better than cure: A cloud incidents perspective.
\newblock In {\em Proceedings of the 31st Joint European Software Engineering
  Conference and Symposium on the Foundations of Software Engineering
  (ESEC/FSE)\/} (2023).

\bibitem{gao2018empirical}
{\sc Gao, Y., Dou, W., Qin, F., Gao, C., Wang, D., Wei, J., Huang, R., Zhou,
  L., and Wu, Y.}
\newblock An empirical study on crash recovery bugs in large-scale distributed
  systems.
\newblock In {\em Proceedings of the 26th ACM joint meeting on european
  software engineering conference and symposium on the foundations of software
  engineering (ESEC/FSE'18)\/} (2018).

\bibitem{ghosh2022fight}
{\sc Ghosh, S., Shetty, M., Bansal, C., and Nath, S.}
\newblock How to fight production incidents? an empirical study on a
  large-scale cloud service.
\newblock In {\em Proceedings of the 13th Symposium on Cloud Computing\/}
  (2022).

\bibitem{gu:sosp:23}
{\sc Gu, J.~T., Sun, X., Zhang, W., Jiang, Y., Wang, C., Vaziri, M., Legunsen,
  O., and Xu, T.}
\newblock {Acto: Automatic End-to-End Testing for Operation Correctness of
  Cloud System Management}.
\newblock In {\em Proceedings of the 29th ACM Symposium on Operating Systems
  Principles (SOSP'23)\/} (2023).

\bibitem{he2022empirical}
{\sc He, S., Zhang, X., He, P., Xu, Y., Li, L., Kang, Y., Ma, M., Wei, Y.,
  Dang, Y., Rajmohan, S., et~al.}
\newblock An empirical study of log analysis at microsoft.
\newblock In {\em Proceedings of the 30th ACM Joint European Software
  Engineering Conference and Symposium on the Foundations of Software
  Engineering (ESEC/FSE)\/} (2022).

\bibitem{inam2022sok}
{\sc Inam, M.~A., Chen, Y., Goyal, A., Liu, J., Mink, J., Michael, N., Gaur,
  S., Bates, A., and Hassan, W.~U.}
\newblock Sok: History is a vast early warning system: Auditing the provenance
  of system intrusions.
\newblock In {\em 2023 IEEE Symposium on Security and Privacy (S\&P'22)\/}
  (2022).

\bibitem{jiang2020mitigate}
{\sc Jiang, J., Lu, W., Chen, J., Lin, Q., Zhao, P., Kang, Y., Zhang, H.,
  Xiong, Y., Gao, F., Xu, Z., et~al.}
\newblock How to mitigate the incident? an effective troubleshooting guide
  recommendation technique for online service systems.
\newblock In {\em Proceedings of the 28th ACM Joint Meeting on European
  Software Engineering Conference and Symposium on the Foundations of Software
  Engineering (ESEC/FSE'20)\/} (2020).

\bibitem{jin2023assess}
{\sc Jin, P., Zhang, S., Ma, M., Li, H., Kang, Y., Li, L., Liu, Y., Qiao, B.,
  Zhang, C., Zhao, P., et~al.}
\newblock Assess and summarize: Improve outage understanding with large
  language models.
\newblock In {\em Proceedings of the Joint European Software Engineering
  Conference and Symposium on the Foundations of Software Engineering
  (ESEC/FSE)\/} (2023).

\bibitem{kasai2023evaluating}
{\sc Kasai, J., Kasai, Y., Sakaguchi, K., Yamada, Y., and Radev, D.}
\newblock Evaluating gpt-4 and chatgpt on japanese medical licensing
  examinations.
\newblock {\em arXiv preprint arXiv:2303.18027\/} (2023).

\bibitem{leesatapornwongsa2017scalability}
{\sc Leesatapornwongsa, T., Stuardo, C.~A., Suminto, R.~O., Ke, H., Lukman,
  J.~F., and Gunawi, H.~S.}
\newblock Scalability bugs: When 100-node testing is not enough.
\newblock In {\em Proceedings of the 16th Workshop on Hot Topics in Operating
  Systems (HotOS'17)\/} (2017).

\bibitem{haozhe2023codec}
{\sc Li, H., Ma, M., Liu, Y., Qin, S., Qiao, B., Yao, R., Chaturvedi, H., Tran,
  T., Chintalapati, M., Rajmohan, S., Lin, Q., and Zhang, D.}
\newblock Codec: Cost-effective duration prediction system for deadline
  scheduling in the cloud.
\newblock In {\em Proceedings of the 34th IEEE International Symposium on
  Software Reliability Engineering\/} (2023).

\bibitem{li2022mining}
{\sc Li, M., Ma, M., Nie, X., Yin, K., Cao, L., Wen, X., Yuan, Z., Wu, D., Li,
  G., Liu, W., et~al.}
\newblock Mining fluctuation propagation graph among time series with active
  learning.
\newblock In {\em Database and Expert Systems Applications: 33rd International
  Conference\/} (2022).

\bibitem{li2021practical}
{\sc Li, Z., Chen, J., Jiao, R., Zhao, N., Wang, Z., Zhang, S., Wu, Y., Jiang,
  L., Yan, L., Wang, Z., et~al.}
\newblock Practical root cause localization for microservice systems via trace
  analysis.
\newblock In {\em 2021 IEEE/ACM 29th International Symposium on Quality of
  Service\/} (2021).

\bibitem{li2023did}
{\sc Li, Z., Luo, C., Chen, T.-H., Shang, W., He, S., Lin, Q., and Zhang, D.}
\newblock Did we miss something important? studying and exploring
  variable-aware log abstraction.
\newblock {\em arXiv preprint arXiv:2304.11391\/} (2023).

\bibitem{lian2023configuration}
{\sc Lian, X., Chen, Y., Cheng, R., Huang, J., Thakkar, P., and Xu, T.}
\newblock Configuration validation with large language models.
\newblock {\em arXiv preprint arXiv:2310.09690\/} (2023).

\bibitem{liu2021microhecl}
{\sc Liu, D., He, C., Peng, X., Lin, F., Zhang, C., Gong, S., Li, Z., Ou, J.,
  and Wu, Z.}
\newblock Microhecl: High-efficient root cause localization in large-scale
  microservice systems.
\newblock In {\em 2021 IEEE/ACM 43rd International Conference on Software
  Engineering: Software Engineering in Practice (ICSE-SEIP'21)\/} (2021).

\bibitem{liu2019bugs}
{\sc Liu, H., Lu, S., Musuvathi, M., and Nath, S.}
\newblock What bugs cause production cloud incidents?
\newblock In {\em Proceedings of the Workshop on Hot Topics in Operating
  Systems (HotOS'19)\/} (2019).

\bibitem{liu2022uniparser}
{\sc Liu, Y., Zhang, X., He, S., Zhang, H., Li, L., Kang, Y., Xu, Y., Ma, M.,
  Lin, Q., Dang, Y., et~al.}
\newblock Uniparser: A unified log parser for heterogeneous log data.
\newblock In {\em Proceedings of the ACM Web Conference 2022\/} (2022).

\bibitem{ResinOSDI2022}
{\sc Lou, C., Chen, C., Huang, P., Dang, Y., Qin, S., Yang, X., Li, X., Lin,
  Q., and Chintalapati, M.}
\newblock {RESIN}: A holistic service for dealing with memory leaks in
  production cloud infrastructure.
\newblock In {\em Proceedings of the 16th USENIX Symposium on Operating Systems
  Design and Implementation (OSDI'22)\/} (2022).

\bibitem{lou2020understanding}
{\sc Lou, C., Huang, P., and Smith, S.}
\newblock Understanding, detecting and localizing partial failures in large
  system software.
\newblock In {\em Proceedings of the 17th USENIX Symposium on Networked Systems
  Design and Implementation (NSDI'20)\/} (2020).

\bibitem{luo2014correlating}
{\sc Luo, C., Lou, J.-G., Lin, Q., Fu, Q., Ding, R., Zhang, D., and Wang, Z.}
\newblock Correlating events with time series for incident diagnosis.
\newblock In {\em Proceedings of the 20th ACM SIGKDD international conference
  on Knowledge discovery and data mining\/} (2014).

\bibitem{ma2020automap}
{\sc Ma, M., Xu, J., Wang, Y., Chen, P., Zhang, Z., and Wang, P.}
\newblock Automap: Diagnose your microservice-based web applications
  automatically.
\newblock In {\em Proceedings of The Web Conference 2020\/} (2020).

\bibitem{ma2020diagnosing}
{\sc Ma, M., Yin, Z., Zhang, S., Wang, S., Zheng, C., Jiang, X., Hu, H., Luo,
  C., Li, Y., Qiu, N., et~al.}
\newblock Diagnosing root causes of intermittent slow queries in cloud
  databases.
\newblock {\em Proceedings of the VLDB Endowment (VLDB'20)\/} (2020).

\bibitem{ma2021jump}
{\sc Ma, M., Zhang, S., Chen, J., Xu, J., Li, H., Lin, Y., Nie, X., Zhou, B.,
  Wang, Y., and Pei, D.}
\newblock Jump-starting multivariate time series anomaly detection for online
  service systems.
\newblock In {\em 2021 USENIX Annual Technical Conference (ATC'21)\/} (2021).

\bibitem{ma2018robust}
{\sc Ma, M., Zhang, S., Pei, D., Huang, X., and Dai, H.}
\newblock Robust and rapid adaption for concept drift in software system
  anomaly detection.
\newblock In {\em 2018 IEEE 29th International Symposium on Software
  Reliability Engineering (ISSRE'18)\/} (2018).

\bibitem{mastropaolo2022using}
{\sc Mastropaolo, A., Pascarella, L., and Bavota, G.}
\newblock Using deep learning to generate complete log statements.
\newblock In {\em Proceedings of the 44th International Conference on Software
  Engineering (ICSE'22)\/} (2022).

\bibitem{mastropaolo2021studying}
{\sc Mastropaolo, A., Scalabrino, S., Cooper, N., Palacio, D.~N., Poshyvanyk,
  D., Oliveto, R., and Bavota, G.}
\newblock Studying the usage of text-to-text transfer transformer to support
  code-related tasks.
\newblock In {\em Proceedings of the 43rd International Conference on Software
  Engineering (ICSE'21)\/} (2021).

\bibitem{tiktoken}
{\sc OpenAI}.
\newblock {Tiktoken}: A python library for tokenizing text.
\newblock \url{https://github.com/openai/tiktoken}, 2023.

\bibitem{shetty2022autotsg}
{\sc Shetty, M., Bansal, C., Upadhyayula, S.~P., Radhakrishna, A., and Gupta,
  A.}
\newblock Autotsg: learning and synthesis for incident troubleshooting.
\newblock In {\em Proceedings of the 30th ACM Joint European Software
  Engineering Conference and Symposium on the Foundations of Software
  Engineering (ESEC/FSE'22)\/} (2022).

\bibitem{sun:osdi:20}
{\sc Sun, X., Cheng, R., Chen, J., Ang, E., Legunsen, O., and Xu, T.}
\newblock {Testing Configuration Changes in Context to Prevent Production
  Failures}.
\newblock In {\em Proceedings of the 14th USENIX Symposium on Operating Systems
  Design and Implementation (OSDI'20)\/} (2020).

\bibitem{sun:osdi:22}
{\sc Sun, X., Luo, W., Gu, J.~T., Ganesan, A., Alagappan, R., Gasch, M.,
  Suresh, L., and Xu, T.}
\newblock {Automatic Reliability Testing for Cluster Management Controllers}.
\newblock In {\em Proceedings of the 16th USENIX Symposium on Operating Systems
  Design and Implementation (OSDI'22)\/} (2022).

\bibitem{tan2023evaluation}
{\sc Tan, Y., Min, D., Li, Y., Li, W., Hu, N., Chen, Y., and Qi, G.}
\newblock Evaluation of chatgpt as a question answering system for answering
  complex questions.
\newblock {\em arXiv preprint arXiv:2303.07992\/} (2023).

\bibitem{wei2022chain}
{\sc Wei, J., Wang, X., Schuurmans, D., Bosma, M., Chi, E., Le, Q., and Zhou,
  D.}
\newblock Chain of thought prompting elicits reasoning in large language
  models.
\newblock {\em arXiv preprint arXiv:2201.11903\/} (2022).

\bibitem{wu2017automated}
{\sc Wu, Y., Chen, A., Haeberlen, A., Zhou, W., and Loo, B.~T.}
\newblock Automated bug removal for software-defined networks.
\newblock In {\em Proceedings of the 14th USENIX Symposium on Networked Systems
  Design and Implementation (NSDI'17)\/} (2017).

\bibitem{xie2023unsupervised}
{\sc Xie, Z., Xu, H., Chen, W., Li, W., Jiang, H., Su, L., Wang, H., and Pei,
  D.}
\newblock Unsupervised anomaly detection on microservice traces through graph
  vae.
\newblock In {\em Proceedings of the ACM Web Conference 2023\/} (2023).

\bibitem{yan2023aegis}
{\sc Yan, X., Hsieh, K., Liyanage, Y., Ma, M., Chintalapati, M., Lin, Q., Dang,
  Y., and Zhang, D.}
\newblock Aegis: Attribution of control plane change impact across layers and
  components for cloud systems.
\newblock In {\em 2023 IEEE/ACM 45th International Conference on Software
  Engineering: Software Engineering in Practice (ICSE-SEIP'23)\/} (2023).

\bibitem{yuan2014simple}
{\sc Yuan, D., Luo, Y., Zhuang, X., Rodrigues, G.~R., Zhao, X., Zhang, Y.,
  Jain, P., and Stumm, M.}
\newblock Simple testing can prevent most critical failures: An analysis of
  production failures in distributed data-intensive systems.
\newblock In {\em Proceedings of the 12th USENIX Symposium on Operating Systems
  Design and Implementation (OSDI'14)\/} (2014).

\bibitem{zeng2021watson}
{\sc Zeng, J., Chua, Z.~L., Chen, Y., Ji, K., Liang, Z., and Mao, J.}
\newblock Watson: Abstracting behaviors from audit logs via aggregation of
  contextual semantics.
\newblock In {\em Network and Distributed System Security Symposium
  (NDSS'21)\/} (2021).

\bibitem{zeng2022shadewatcher}
{\sc Zeng, J., Wang, X., Liu, J., Chen, Y., Liang, Z., Chua, T.-S., and Chua,
  Z.~L.}
\newblock Shadewatcher: Recommendation-guided cyber threat analysis using
  system audit records.
\newblock In {\em 2022 IEEE Symposium on Security and Privacy (S\&P'22)\/}
  (2022).

\bibitem{zeng2023traceark}
{\sc Zeng, Z., Zhang, Y., Xu, Y., Ma, M., Qiao, B., Zou, W., Chen, Q., Zhang,
  M., Zhang, X., Zhang, H., et~al.}
\newblock Traceark: Towards actionable performance anomaly alerting for online
  service systems.
\newblock In {\em 2023 IEEE/ACM 45th International Conference on Software
  Engineering: Software Engineering in Practice (ICSE-SEIP'23)\/} (2023).

\bibitem{TraceArk}
{\sc Zeng, Z., Zhang, Y., Xu, Y., Ma, M., Qiao, B., Zou, W., Chen, Q., Zhang,
  M., Zhang, X., Zhang, H., Gao, X., Fan, H., Rajmohan, S., Lin, Q., and Zhang,
  D.}
\newblock Traceark: Towards actionable performance anomaly alerting for online
  service systems.
\newblock In {\em To appear in Proc. of ICSE\/} (2023).

\bibitem{zhang2022using}
{\sc Zhang, J., Mytkowicz, T., Kaufman, M., Piskac, R., and Lahiri, S.~K.}
\newblock Using pre-trained language models to resolve textual and semantic
  merge conflicts (experience paper).
\newblock In {\em Proceedings of the 31st ACM SIGSOFT International Symposium
  on Software Testing and Analysis\/} (2022).

\bibitem{zhang2023system}
{\sc Zhang, T., Qiu, H., Castellano, G., Rifai, M., Chen, C.~S., and Pianese,
  F.}
\newblock System log parsing: A survey.
\newblock {\em IEEE Transactions on Knowledge and Data Engineering\/} (2023).

\bibitem{zhang2021onion}
{\sc Zhang, X., Xu, Y., Qin, S., He, S., Qiao, B., Li, Z., Zhang, H., Li, X.,
  Dang, Y., Lin, Q., et~al.}
\newblock Onion: identifying incident-indicating logs for cloud systems.
\newblock In {\em Proceedings of the 29th ACM Joint Meeting on European
  Software Engineering Conference and Symposium on the Foundations of Software
  Engineering\/} (2021).

\bibitem{zhang2021cloudrca}
{\sc Zhang, Y., Guan, Z., Qian, H., Xu, L., Liu, H., Wen, Q., Sun, L., Jiang,
  J., Fan, L., and Ke, M.}
\newblock Cloudrca: a root cause analysis framework for cloud computing
  platforms.
\newblock In {\em Proceedings of the 30th ACM International Conference on
  Information \& Knowledge Management\/} (2021).

\bibitem{zhang2021understanding}
{\sc Zhang, Y., Yang, J., Jin, Z., Sethi, U., Rodrigues, K., Lu, S., and Yuan,
  D.}
\newblock Understanding and detecting software upgrade failures in distributed
  systems.
\newblock In {\em Proceedings of the ACM SIGOPS 28th Symposium on Operating
  Systems Principles (SOSP'21)\/} (2021).

\bibitem{zhang2023automatic}
{\sc Zhang, Z., Zhang, A., Li, M., and Smola, A.}
\newblock Automatic chain of thought prompting in large language models.
\newblock In {\em The Eleventh International Conference on Learning
  Representations (ICLR'23)\/} (2023).

\bibitem{zhao2023robust}
{\sc Zhao, C., Ma, M., Zhong, Z., Zhang, S., Tan, Z., Xiong, X., Yu, L., Feng,
  J., Sun, Y., Zhang, Y., et~al.}
\newblock Robust multimodal failure detection for microservice systems.
\newblock In {\em Proceedings of the 29th ACM SIGKDD Conference on Knowledge
  Discovery and Data Mining\/} (2023).

\end{thebibliography}

\clearpage
\label{sec:conclusion}

\end{document}